\documentclass{article}
\usepackage{PRIMEarxiv}
\usepackage{apacite}
\usepackage{natbib}
\pdfoutput=1

\usepackage[utf8]{inputenc} 
\usepackage[T1]{fontenc}    
\usepackage{url}            
\usepackage{booktabs}       
\usepackage{amsfonts}       
\usepackage{nicefrac}       
\usepackage{microtype}      
\usepackage{lipsum}
\usepackage{fancyhdr}       
\usepackage{graphicx}       
\graphicspath{{media/}}     
\usepackage{xcolor}
\usepackage{amsmath,amssymb}
\usepackage{listings}
\usepackage{tabularx}
\usepackage{color}

\title{Modelling wildfire spread and spotfire merger using conformal mapping and AAA-least squares methods}

\author{
  Samuel J. Harris and N.R. McDonald \\
  Department of Mathematics \\ 
  University College London \\
  Gower Street, London, WC1E 6BT, UK \\
  sam.harris.16@ucl.ac.uk 
}

\begin{document}

\maketitle

\begin{abstract}
A two-dimensional model of wildfire spread and merger is presented. Three features affect the fire propagation: (i) a constant basic rate of spread term accounting for radiative and convective heat transfer, (ii) the unidirectional, constant ambient wind, and (iii) a fire-induced pyrogenic wind. Two numerical methods are proposed to solve for the harmonic pyrogenic potential. The first utilises the conformal invariance of Laplace’s equation, reducing the wildfire system to a single Polubarinova-Galin type equation. The second method uses a AAA-least squares method to find a rational approximation of the potential. Various wildfire scenarios are presented and the effects of the pyrogenic wind and the radiative/convective basic rate of spread terms investigated. Firebreaks such as roads and lakes are also included and solutions are found to match well with existing numerical and experimental results. The methods proposed in this work are suitably fast and accurate to be considered for operational use.
\end{abstract}


\section{Introduction}\label{sec:intro}
Climate change predictions indicate a continued, substantial shift in global weather conditions. Fire weather conditions, as defined in e.g. \cite{jolly2015climate}, are forecast to increase worldwide over the coming decades which would result in an enhanced level of ``fire danger'' and an increased risk of catastrophic wildfire development \citep{arnell2021effect,IPCC,hetzer2024fire}. The growing prevalence of wildfires is now a global phenomenon with large, dynamic wildfires - so called mega fires - increasing in frequency and intensity worldwide \citep{attiwell2013exploring}. Wildfire research, specifically that involving the complex and nonlinear dynamics and thermodynamics associated with wildland fire spread, is consequently receiving growing attention.

There are many active branches of wildfire study, each presenting their own persisting challenges. These include (but are not limited to): the combustion of different fuel types \citep{santoni2014bulk,liu2021combustion,sullivan2022wildland}; distribution of fuels in a fuel bed \citep{catchpole1989fire,khan2023simulated}; investigation of terrain influence on fire spread \citep{malangone2011effects,ambroz2019numerical,hilton2021rapid,ribeiro2023slope}; generation and spread of spotfires \citep{boychuk2009stochastic,martin2016spotting,bonta2017intentional,urban2019ignition}; deployment and optimisation of fire management and suppression efforts \citep{izhaki1997effects,hu2007agent,ausonio2021drone,yfantidou2023geoinformatics} and; the coupled interaction of the wildfire with the atmosphere \citep{clark1996bcoupled,sun2009importance,mandel2011coupled,bakhshaii}. Dynamic weather conditions such as strong, volatile  winds cause equally dynamic and often unpredictable fire spread \citep{viegas2012study,wheeler,thomas2017modelling}. Further, the fire itself affects the overall wind field: the wildfire generates its own wind source known as the pyrogenic (or fire) wind \citep{smith1975role,beer,hiltonpyro,quaife_speer} as air is drawn toward the fire from the surrounding region and ejected vertically from the burnt region via a ``fire plume''. How this cumulative wind flow affects the fire spread itself is a focus of this work. 

Additionally, this work seeks to develop and demonstrate a model capable of operational use. There are numerous existing wildfire models that produce accurate results as well as those which incorporate a full three-dimensional wildfire-atmosphere coupling - see the recent review articles by \cite{papadopoulos2011comparative} and \cite{bakhshaii}. Such models are useful in certain situations, for example if the user wanted to study a previous fire scenario in-depth or wanted to assess the wildfire spread of a hypothetical future fire event. However, the complexity and high computation cost of these models (which can be on the order of days) means that they cannot be used to predict future wildfire spread of a current fire event in real time with a given fire line geometry and weather conditions. In these cases, simpler and computationally cheaper models are more suitable as they can run ``faster than real time'': on the order of hours, minutes or seconds. The challenge is then to make these models sufficiently accurate to, for example,  give a realistic prediction of wildfire spread.

One common simplification is to reduce the fully three-dimensional problem to two-dimensions, and turn the focus to modelling the spread of the wildfire on a two-dimensional surface. This is particularly suitable when modelling surface fires - fires with fuel elements under 1.8m tall \citep{mell2007physics} - which are exclusively considered in this work. The work of \cite{quaife_speer} includes the wildfire-atmosphere interaction in their cellular automata model, in contrast to this work which uses a continuum model. In such two-dimensional models, the wildfire is represented by the fire perimeter or ``fire line'': a simple, singly connected curve separating the burnt and unburnt regions of fuel. The evolution of the fire line is then sought. 

Over nearly the past decade, Hilton, Sharples and their colleagues have successfully developed such a two-dimensional continuum model of surface fire spread which incorporates a dynamic wildfire-atmosphere interaction, see for example \cite{hiltonpyro, sharples2020modeling} and \cite{sharples2022fire}. This model has been shown to perform well, demonstrating good agreement between experimental wildfire data on small and large scales, all while running quickly and efficiently. Taking inspiration from this success, this work similarly aspires to formulate and solve a simple  two-dimensional continuum model of wildfire spread. As in \cite{hiltonpyro}, it is assumed that the wildfire spreads on flat terrain through a single type, homogeneously distributed surface fuel and thus the fire evolution is affected by two factors. The first is a basic rate of spread (ROS) term: this is some (known) constant which captures the physical, chemical and thermodynamic processes of combustion and heat transfer for a given fuel element. The second is the surrounding wind field: this is in turn composed of some constant, unidirectional ambient wind and the self-induced pyrogenic wind - the latter of which must be solved for in the region exterior to the fire line. While the model of this work is closely related to that of \cite{hiltonpyro}, there are three notable differences: (a) the basic ROS term is explicitly split into its radiative and convective components as the convective term will interact directly with the wind field; the radiative part will not; (b) the fire plume over the burnt region of fuel is here observed at infinity as an effective point sink of air with strength $Q$; (c) whereas \cite{hiltonpyro} use the Spark framework \citep{miller2015spark} to compute the evolution of the fire line which is based on the level set method \citep{osher1988fronts, mallet2009modeling}, two alternative numerical methods are proposed in this work.

The first numerical procedure is based on a conformal mapping method and is used exclusively for single wildfire problems. Conformal mapping is the  transforming of a two-dimensional domain in the ``physical'' plane to a corresponding domain in another ``canonical'' plane where the angles between points are conserved under the mapping \citep{brown2009complex}. By the Riemann mapping theorem, such a map between domains always exists and thus the exterior fire region (and the fire line itself) can be transformed to the desired canonical domain - here, the exterior of the unit disk - for all time as the fire line evolves. Further, the associated governing equation (the Laplace equation) and Dirichlet boundary condition are unchanged under conformal mapping and thus the pyrogenic potential can be solved exactly in the canonical domain. Wind and other dynamical effects on the wildfire can then be incorporated into   a single equation of Polubarinova-Galin (PG) type \citep{harris2022fingering} for the time-dependent conformal map $f$ from the disk exterior to the fire line exterior. This is a common technique used in solving similar free boundary problems with conformally invariant governing equations, for example in fluid dynamics and mathematical biology \citep{entov1991bubble,bazant2005conformal,ladd2020,harris2023penguin}. 

Now, the problem is to find the conformal map $f(\zeta,t)$ from the unit disk to the fire line at time $t$. For all time, this can be approximated as a truncated power series in $\zeta$ - the complex coordinate in the canonical plane - plus the initial conformal map $f(\zeta,0)=g(\zeta)$. This is found numerically by means of the Schwarz-Christoffel (SC) Toolbox: a  MATLAB package developed in \cite{driscoll2002schwarz,driscoll2005algorithm} for finding conformal maps between polygons and various canonical domains, including the unit disk. The conformal map of some special types of shape (known in this work as ``Laurent shapes'') can immediately be written as a truncated Laurent series, thus mitigating the need for the SC Toolbox in these cases. Thus the problem reduces to solving a system of $n$ ODEs in time which can be done simply and efficiently using any standard numerical ODE solver: MATLAB's \textit{ode15i} is used in this work.

The second numerical method is able to handle multiple interacting wildfires which may have started independently or may be a configuration of multiple spotfires resulting from one main wildfire. These fires interact via each fire's perturbation of the wind field, in particular it is observed that two fires will grow towards each other until the fires eventually merge \citep{hiltonpyro} since this region has a reduced pyrogenic wind strength. A conformal mapping method is difficult to implement in the multiple fire scenario: there is no simple form for the conformal map of this multiply connected domain to some desirable canonical domain (which is no longer the simple unit disk exterior) that is valid for all time, especially if some (or all) of the fires merge and thus change the connectedness of the domain at some time $t>0$. 

Instead, the pyrogenic potential is solved directly in the physical plane via a rational approximation of the potential which can then be found numerically using a combination of the adaptive Antoulas Anderson (AAA, pronounced ``triple A'') algorithm and a least-squares (LS) method. This combined ``AAA-LS'' algorithm has been developed by Trefethen, Costa, Gopal and co-workers over the past five years \citep{nakatsukasa2023first} and has been applied to a variety of problems in fluid dynamics and electrostatics, see for example \cite{trefethen2018series,trefethen2020numerical,nakatsukasa2018aaa,gopal2019solving,costa2023aaa}. The approximation of the potential is composed of a polynomial and a sum of rational terms involving simple poles in the unphysical interior domains. The AAA-LS method calculates these poles and the unknown (complex) coefficients using boundary data and the corresponding Dirichlet boundary condition for the potential. This algorithm is  fast, running in seconds on a standard laptop, and boasts root-exponential convergence with respect to the number of simple poles used \citep{gopal2019solving}. An explicit Runge-Kutta time stepping method \citep{butcher1996history} can then be implemented to track the fire line progression over time with the simple MATLAB function \textit{union} used to detect and merge any overlapping fire lines.

Both proposed methods offer their own merits. The appeal of the single wildfire conformal mapping method is that it uses an implicit ODE solver and so runs quickly, even for many time steps. The multiple fire method requires explicit time stepping, so even though one iteration of the AAA-LS algorithm runs quickly, hundreds of time steps may be desired which then adds to the computation time. However, this method can be extended to handle more complicated wildfire scenarios than the conformal mapping method can, for example those including multiple wildfires, fire merger, firebreaks and point vortices representing fire whirls (not explored in this work). Therefore, both methods are presented in this work for comparison.

This paper is laid out as follows. Sections \ref{sec:modelsing} and \ref{sec:modelmult} describe the model setup for the single and multiple wildfire scenarios, respectively. Section \ref{sec:cmap} outlines the single wildfire conformal mapping method and section \ref{sec:aaa} the AAA-least squares algorithm and multiple spotfires method. Some examples of single and multiple wildfires are then presented in sections \ref{sec:resultssing} and \ref{sec:resultsmult}, respectively, showing the effects that pyrogenic wind and the radiative and convective basic ROS have on the development of the fire line. Firebreaks such as roads, rivers and lakes are included into the multiple fire model in section \ref{sec:fbreak}. Then in section \ref{sec:hcomp}, the model of this work is compared against the pyrogenic potential model of \cite{hiltonpyro} and the experimental results produced in \cite{sullivan2019investigation}. A final discussion is given in section \ref{sec:conc} which highlights areas of extension for the model presented in this work. 

\section{Single wildfire model}\label{sec:modelsing}
\begin{figure}[ht]
\centering
\includegraphics[width=0.99\textwidth]{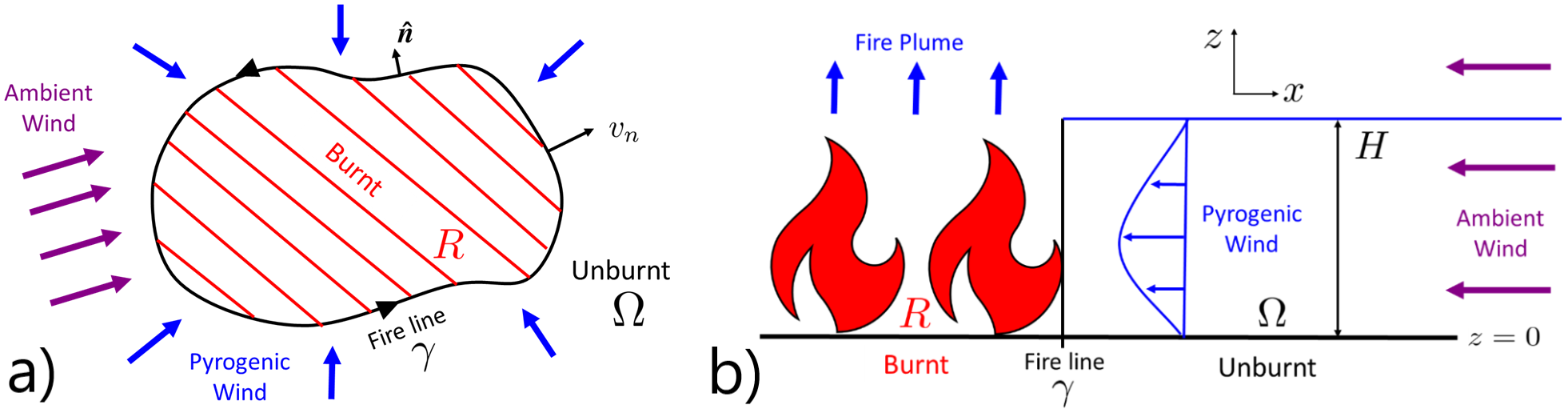}
\caption{The single wildfire model (a) plan view, (b) side view. The blue arrows represent the direction of the pyrogenic (fire) wind and the purple arrows the direction of an ambient wind.}\label{fig:singlefire}
\end{figure}

Consider a single wildfire spreading across a flat bed of homogeneously distributed, single-type fuel.  Surface fires are considered here, i.e. the fire spreads through surface fuels such as grasses, heather, fallen leaves and twigs and shrubs less than 1.8m tall \citep{mell2007physics}. This implies that the horizontal length scale of the fire is suitably larger than the vertical flame height $H$ and thus the problem can be treated as two-dimensional; this is illustrated in figure \ref{fig:singlefire}. 

The fire line (or fire perimeter) denoted by $\gamma$ is the finite, closed Jordan curve separating the two distinct regions of singly-connected, finite-area burnt $R$ and unburnt $\Omega$ fuels. Modelling the wildfire spread becomes that of finding the evolution of this curve when subject to various dynamical effects. In practice, this amounts to solving for the velocity of the fire line in the normal direction $\boldsymbol{\hat{n}}$ - tangential velocities do not affect the outward spread of the fire. In this work, three key factors are considered which affect the normal velocity $v_n$. 

First, the basic rate of spread (ROS) term $v_0$ captures the physical, chemical and thermodynamic processes involved in fuel ignition and heat transfer. In the absence of all other external factors such as wind and terrain, the fire spreads at this constant speed $v_0$ in the direction of the local normal to the fire line. The exact value of this ROS term for different fuels is still an open question, see the review articles by e.g. \cite{sull2009a} and \cite{morvan2022fifty}, due to a continued lack of understanding of both the ignition process and of the relative magnitudes of convection and radiation in heat transfer. Others have proposed models in which the ROS is calculated deterministically  rather than given as an ad-hoc constant \citep{mcdonald2024exact,dipierro2024simple}. In this work, an analysis into neither the value of $v_0$ nor the quantity $\alpha\in[0,1]$ representing the proportion of ROS relating to radiative effects (so that $(1-\alpha)$ represents the convective effect) is performed here. Instead, the basic ROS is simply taken as some constant $v_0 = \alpha v_0 + (1-\alpha)v_0$ for a given $\alpha$. This explicit distinction between radiative and convective terms\footnote{It is assumed that conductive effects are sufficiently weak that they can be neglected.} is similar to that of \cite{beer} and the constant $\alpha$ is almost an exact analogy of the dimensionless parameter $P$ defined in \cite{baines1990physical}.

Second, there is a unidirectional, constant strength ambient wind $\boldsymbol{u}_a=U_a\boldsymbol{\hat{u}}_a$ present, of magnitude $U_a$ and dimensionless, unit direction $\boldsymbol{\hat{u}}_a$. It is assumed that the fire line $\gamma$ is entirely permeable to this ambient wind, thus all sections of the fire line experience the same (constant) ambient wind effect. The side of the fire line facing the oncoming ambient wind is labelled as the windward side and the opposite side as the leeward side. Third, the fire itself generates a pyrogenic (or fire) wind $\boldsymbol{u}_p$ \citep{trelles1997fire,lareau,thomas2017modelling,hiltonpyro,sullivan2019investigation}. Air in the burnt region $R$ is ejected vertically  out of the two-dimensional system in figure \ref{fig:singlefire}. The resulting low pressure at the fire line induces a horizontal (dynamic) pressure gradient, driving a pyrogenic wind towards the fire line where it too is ejected as part of the fire plume. The overall wind vector can then simply be written as the linear combination of the ambient and pyrogenic winds $\boldsymbol{u}=\boldsymbol{u}_p+\boldsymbol{u}_a$.  

Note that the ROS and ambient wind $\boldsymbol{u}_a$ have purely kinematic effects on the fire line evolution. In contrast, the pyrogenic wind is a dynamical effect since the shape of the fire line determines $\boldsymbol{u}_p$. Thus the co-evolution of    $\boldsymbol{u}_p$ and the fire line represent a nonlinear, two-dimensional free boundary problem.

Some simplifying assumptions are made. First, the ambient wind is such that it is not deviated by the fire plume (the permeability assumption mentioned above). Second, any (vertical) deviations of the fire plume induced by the ambient wind do not affect the fire spread at the horizontal, surface plane. Third, the fire plume acts as a sink on the surface drawing in surrounding air. The strength $Q$ of the fire plume, the effective strength of the sink, is assumed constant for all time. An extension of the model involving a time varying plume strength as the fire grows is of interest and has been explored in previous work - see \cite{harris2022fingering}.

While the basic ROS and ambient wind are given constants, the pyrogenic wind must be solved in the fire line exterior $\Omega$ for the evolving fire line $\gamma$. To good approximation, the pyrogenic wind can be treated as an irrotational flow of an incompressible fluid thus $\boldsymbol{u}_p=\boldsymbol{\nabla}\phi$ where the velocity potential $\phi$ satisfies the Laplace equation \citep{hiltonpyro,sharples2020modeling,quaife_speer,harris2022fingering}
\begin{equation}
\nabla^2\phi=0\;\;\;\text{in }\Omega.
\end{equation}
The pyrogenic wind is assumed to flow horizontally in a shallow layer of depth $H$ at ground level - the same as the flame height in figure \ref{fig:singlefire} - with $\boldsymbol{u}_p$ the average wind velocity over $H$. Consider its Reynolds number Re $=UH/D$, where $U$ is some representative speed of the pyrogenic wind and $D$ is its horizontal (momentum) diffusivity. From figure 6 of \cite{bebieva} and table III of \cite{beer} using an ambient wind speed of $U_a=1$ms$^{-1}$ (a representative value used throughout this work), it follows that $U=0.5$ms$^{-1}$, $H=0.5$m and $D=1$m$^2$s$^{-1}$ and so Re $\approx 0.25$. Thus the shallow pyrogenic wind flow is of sufficiently low Reynolds number here such that the Navier-Stokes equations reduce to Stokes flow. Integrating across the fluid layer of depth $H$ (see e.g. \cite{gustafsson2006conformal}) it follows that $\boldsymbol{u}_p=\boldsymbol{\nabla}\phi\sim-\boldsymbol{\nabla}p$. Thus the velocity potential $\phi$ is proportional to the negative pressure $-p$.

Dynamic pressure best describes the cause of pyrogenic wind flow rather than the simple entrainment laws often used in plume dynamics \citep{smith1975role}. The low pressure at the fire line (caused by the fire plume) is modelled by the boundary condition $p=\phi=0$ on $\gamma$ without loss of generality. In the far field, the fire plume is observed as an effective sink of strength $Q$ and thus $\phi\rightarrow -(Q/2\pi)\log{r}$ as $r\rightarrow\infty$. Note that \cite{hiltonpyro} treat the fire plume differently, instead solving an additional Poisson equation in the burnt region $R$ dependent on the upwards air flow $\nu=-w_z$. As shown in section \ref{sec:hcomp}, the formulation of the pyrogenic wind used here gives similar quantitative results and matching with experimental data as in \cite{hiltonpyro}. 

Finally, the equation for $v_n$, the normal velocity of the fire line $\gamma$, can be formulated. This has the same form as in \cite{hiltonpyro}, namely
\begin{equation}\label{eq:rcopa11}
A v_n = v_0+\tilde{\beta}\boldsymbol{\hat{n}\cdot \nabla}\phi+\tilde{\lambda}\boldsymbol{\hat{n}\cdot u}_a,
\end{equation}
where $\tilde{\beta}$, $\tilde{\lambda}$ are dimensional constants and $A$ is a non-dimensional constant found \textit{a posteriori} when comparing the wildfire propagation with experimental data - see section \ref{sec:hcomp}. One final modification must be made to \eqref{eq:rcopa11} to ensure the fire line satisfies the entropy condition \citep{sethian1985curvature} which states that the fire line cannot intersect previous iterations of itself: once a fuel element is burnt, it cannot become unburnt ie $v_n\geq0$. The radiative proportion of the basic ROS $\alpha v_0$ satisfies this automatically. However, the convective effects of the basic ROS $(1-\alpha)v_0$, the pyrogenic wind $\boldsymbol{u}_p$ and the ambient wind $\boldsymbol{u}_a$ may be large and negative, especially on the windward side of the fire. This wind may be sufficiently strong to stop fire spread by convection (not radiation) but not so strong as to push the fire line backwards, which is nonphysical. Therefore, \eqref{eq:rcopa11} is modified as
\begin{equation}\label{eq:rcopa1}
    A v_n = \alpha v_0 +\max[0,(1-\alpha)v_0+\tilde{\beta}\boldsymbol{\hat{n}\cdot \nabla}\phi+\tilde{\lambda}\boldsymbol{\hat{n}\cdot u}_a],
\end{equation}
ensuring that convective (wind) terms have a non-negative effect on the fire line velocity. 

\subsection{Non-dimensionalisation}\label{subsec:nondim}
Non-dimensional quantities are introduced by scaling with respect to a characteristic velocity and length: the basic ROS velocity $v_0$ and the initial wildfire radius $R_0=R(0)$. In particular, $R_0$ is the conformal radius \citep{bazant2005conformal} in conjunction with the conformal mapping method in section \ref{sec:cmap}. The pyrogenic potential is also scaled by $Q$, the (constant) strength of the fire plume. The resulting (starred) dimensionless variables are 
\begin{equation}\label{eq:scalings}
    \boldsymbol{x} =R_0\boldsymbol{x}^\ast, \;\;\; t = \frac{AR_0}{v_0}t^\ast, \;\;\;
    \boldsymbol{\nabla} =\frac{1}{R_0}\boldsymbol{\nabla}^\ast,  \;\;\; \phi=\frac{Q}{2\pi}\phi^\ast.
\end{equation}
Dropping stars immediately, the normal velocity equation \eqref{eq:rcopa1} becomes
\begin{equation}\label{eq:rcopa2}
v_n = \alpha +\max[0,\;(1-\alpha)+\beta\boldsymbol{\hat{n}\cdot \nabla}\phi+\lambda\boldsymbol{\hat{n}\cdot u}_a].
\end{equation}
where the new dimensionless parameters are 
\begin{equation}\label{eq:params}
    \beta= \frac{\tilde{\beta}\big(\frac{Q}{2\pi R_0}\big)}{v_0}= \frac{\tilde{\beta}U_p}{v_0}, \;\;\; 
    \lambda= \frac{\tilde{\lambda}U_a}{v_0}. 
\end{equation}
Note the labelling $U_p=Q/2\pi R_0$: this is the dimensional magnitude of the pyrogenic wind. The full non-dimensional system governing the motion of the fire line $\gamma$ is thus given by
\begin{gather}
v_n=\alpha + \max[0,\;(1-\alpha) +\beta\boldsymbol{\hat{n}\cdot \nabla}\phi+\lambda\boldsymbol{\hat{n}\cdot \hat{u}}_a]\;\;\;\text{on }\gamma, \label{Meq:rcopa}\\
\nabla^2\phi=0\;\;\;\text{in }\Omega,\label{Meq:Lap}\\
\phi=0\;\;\;\text{on }\gamma,\label{Meq:bdry}\\
\phi\rightarrow-\log{r}\;\;\;\text{as }r\rightarrow\infty.\label{Meq:far}
\end{gather}

Finally, consider the relative magnitudes of the parameters in \eqref{Meq:rcopa} by comparing with the values used in \cite{hiltonpyro} and observations made in \cite{beer}. Wind is the dominant effect and so $\beta,\lambda=\mathcal{O}(10)$, though some $\mathcal{O}(1)$ values may be used in this work for illustrative purposes. By comparing the values of fire and ambient (``mid-flame'') winds from Table III of \cite{beer}, it is assumed that $\lambda\approx2\beta$. Similarly from their Table II, comparing convective and non-convective (radiative) heat fluxes in the absence of an ambient wind gives an approximate value of $\alpha\approx0.5$. Unless stated otherwise, these parameter values will be used throughout this work.

\section{Conformal mapping numerical method}\label{sec:cmap}
The two-dimensional nature of the problem means that it can be formulated in the complex $z=x+iy$ plane which immediately permits the use of methods based on complex analysis. One approach used here involves conformal mapping to describe the wildfire spread. By the Riemann mapping theorem (see e.g. \cite{brown2009complex}), there always exists a  conformal map $z=f(\zeta,t)$ from the exterior of the unit disk in some canonical $\zeta-$plane to the exterior of the fire line $\gamma$ at time $t$ in the physical $z-$plane. The main advantage of conformal mapping in this problem is that  Laplace's equation \eqref{Meq:Lap} is conformally invariant - it is unchanged in this transformation - and so the pyrogenic potential $\phi$ can be solved exactly in the exterior to the $\zeta-$disk
\begin{equation}\label{circlesolns}
    \phi(\zeta) = -\log{|\zeta|}.
\end{equation}
The fire line velocity \eqref{Meq:rcopa} can then be written in terms of this solution and the conformal map $z=f(\zeta,t)$, reducing the system \eqref{Meq:rcopa} - \eqref{Meq:far} in the physical $z-$plane to a single equation of Polubarinova-Galin (PG) type. This approach is commonly used in solving other free boundary problems, for example in \cite{goldstein1978effect, bazant2005conformal,grodzki2019reactive,ladd2020}.

To formulate the PG equation, note the following relations \citep{dallaston2013accurate}
\begin{equation}\label{vnkappan}
v_n=\frac{\text{Re}\Big(f_t\overline{\zeta f_\zeta}\Big)}{|f_\zeta|}, \;\;\;\;\;\; n =\frac{\zeta f_\zeta}{|f_\zeta|},
\end{equation}
where $n=n_x+in_y$ is the complex representation of the outward unit normal vector in the $z-$plane. The pyrogenic wind contribution in \eqref{Meq:rcopa} can be written as
\begin{equation}
\boldsymbol{\hat{n}\cdot\nabla}\phi = \frac{\partial\phi}{\partial n} = \text{Re}[n\overline{\nabla}\phi]=2\text{Re}\bigg[n\frac{\partial \phi}{\partial z}\bigg]=-\frac{1}{|f_\zeta|}\;\;\;\;\text{on }|\zeta|=1, \label{dphi/dnpyro}
\end{equation}
where $\overline{\nabla}=\partial_x -i\partial_y$. This uses the exact solution \eqref{circlesolns} and the equivalence $|\zeta| = (\zeta\overline{\zeta})^{1/2}$ such that $\partial \phi/\partial\zeta=-1/(2\zeta)$ on the unit $\zeta-$disk $|\zeta|=1$. Similarly, the ambient wind contribution is
\begin{equation}\label{ambipyro}
\boldsymbol{\hat{n}\cdot\hat{u}}_a = \text{Re}[n\;\overline{u_a}]=\frac{1}{|f_\zeta|}\text{Re}\Big[\zeta f_{\zeta}\overline{u_a}\Big]\;\;\;\;\text{on }|\zeta|=1,
\end{equation}
where $u_a$ is the unit ambient wind vector written in complex notation. Thus the PG-type equation is
\begin{equation}\label{Meq:PG}
\text{Re}\Big(f_t\overline{\zeta f_\zeta}\Big)=\alpha|f_\zeta| + \max\Big[0,\;(1-\alpha)|f_\zeta| -\beta+\lambda\text{Re}\big(\zeta f_\zeta\overline{u_a}\big)\Big].
\end{equation}

The numerical task is now to find the conformal map $f(\zeta,t)$ from the unit $\zeta-$disk to the fire line $\gamma$. For some special fire line geometries such as circles, ellipses and other n-fold symmetric shapes \citep{dallaston2016curve}, the map $f$ can be written as the following finite Laurent series,
\begin{equation}\label{Meq:LAcmap}
    z = f(\zeta,t) = a_{-1}(t)\zeta + \sum_{k=0}^N c_k(t)\zeta^{-k}.
\end{equation}
These are referred to in this work as Laurent shapes where $a_{-1}(0)$ is the initial conformal radius $R_0$ of the wildfire as defined in section \ref{subsec:nondim}. The series is truncated at $N$ terms for numerical purposes to give $n=2N+3$ unknown real functions in time: $a_{-1}(t)$ and the real and imaginary components of the complex functions $c_k(t)=a_k(t)+ib_k(t)$. Selecting $n$ points around the unit $\zeta-$disk thus converts the PG equation \eqref{Meq:PG} into a system of $n$ ODEs in time $t$ for the $n$ unknown functions, which can be solved using any suitable ODE solver; the MATLAB function $ode15i$ is used in this work.

For general (non-Laurent) shapes, the approximation \eqref{Meq:LAcmap} may be poor for finite $N$ and so the conformal map must be found another way. One approach is to find the map explicitly at each time step, for example by using the SC Toolbox \citep{driscoll2002schwarz,driscoll2005algorithm}. Such an explicit procedure would be computationally expensive and possibly unstable without higher order time stepping procedures, see section \ref{sec:aaa}. 

An implicit method can be implemented by combining the approximation \eqref{Meq:LAcmap} with a single run of the SC Toolbox to compute the initial conformal map $z=f(\zeta,0)=g(\zeta)$. The general map $f(\zeta,t)$ can then be written as
\begin{equation}\label{Meq:SCcmap}
    z = f(\zeta,t) = a_{-1}(t)g(\zeta) + \sum_{k=0}^N c_k(t)\zeta^{k},
\end{equation}
with $a_{-1}(0)=1$ and $c_k(0)=0$, $\forall k$. The initial conformal radius $R_0$ of these shapes is now $a_{-1}(0)A$ where $A$ is the conformal radius of the map $g$. The powers of $\zeta$ are of opposite sign to \eqref{Meq:LAcmap} as the SC map $g(\zeta)$ transforms the interior of the unit disk to the exterior of the fire line. When simulating non-Laurent fire line evolution, the PG equation \eqref{Meq:PG} is modified to account for the conformal map from interior to exterior; the numerical method is otherwise unchanged from that of Laurent shapes.

\subsection{Rate of change of area law}\label{subsec:RCA}
An expression for the rate of change of the area $A(t)$ enclosed by the fire line $\gamma(t)$ can be derived, referred to as the rate of change of area (RCA) law. The RCA law can then be used to check the upcoming numerical results. By integrating the velocity $v_n$ over the curve $\gamma$ and substituting in \eqref{Meq:rcopa}, the rate of change of area is given by
\begin{equation}
    \frac{\mathrm{d}A}{\mathrm{d}t}=\int_{\gamma}v_n\mathrm{d}s = \alpha\int_\gamma \mathrm{d}s + \int_\gamma \max[0,\;(1-\alpha) +\beta\boldsymbol{\hat{n}\cdot \nabla}\phi+\lambda\boldsymbol{\hat{n}\cdot \hat{u}}_a] \mathrm{d}s.
\end{equation}
By definition, the first term on the right hand side is the length $L(t)$ of the fire line curve $\gamma$. The second term is converted to the $\zeta-$plane by integrating along the unit disk $\zeta=e^{i\theta}$ for $\theta\in[0,2\pi)$, thus $\mathrm{d} s = |f_{\zeta}|\mathrm{d}\theta$ where $f$ is the conformal map \eqref{Meq:LAcmap}. Using the expressions \eqref{dphi/dnpyro} and \eqref{ambipyro} gives the RCA law
\begin{equation}\label{eq:RCA}
    \frac{\mathrm{d}A}{\mathrm{d}t}= \alpha L(t)+\int_0^{2\pi} \max\Big[0,\;(1-\alpha)|f_\zeta| -\beta+\lambda\text{Re}\big[\zeta f_\zeta\overline{u_a}\big]\Big] \mathrm{d}\theta.
\end{equation}

\section{Single wildfire results}\label{sec:resultssing}

\begin{figure}[ht]
\centering
\includegraphics[width=0.95\textwidth]{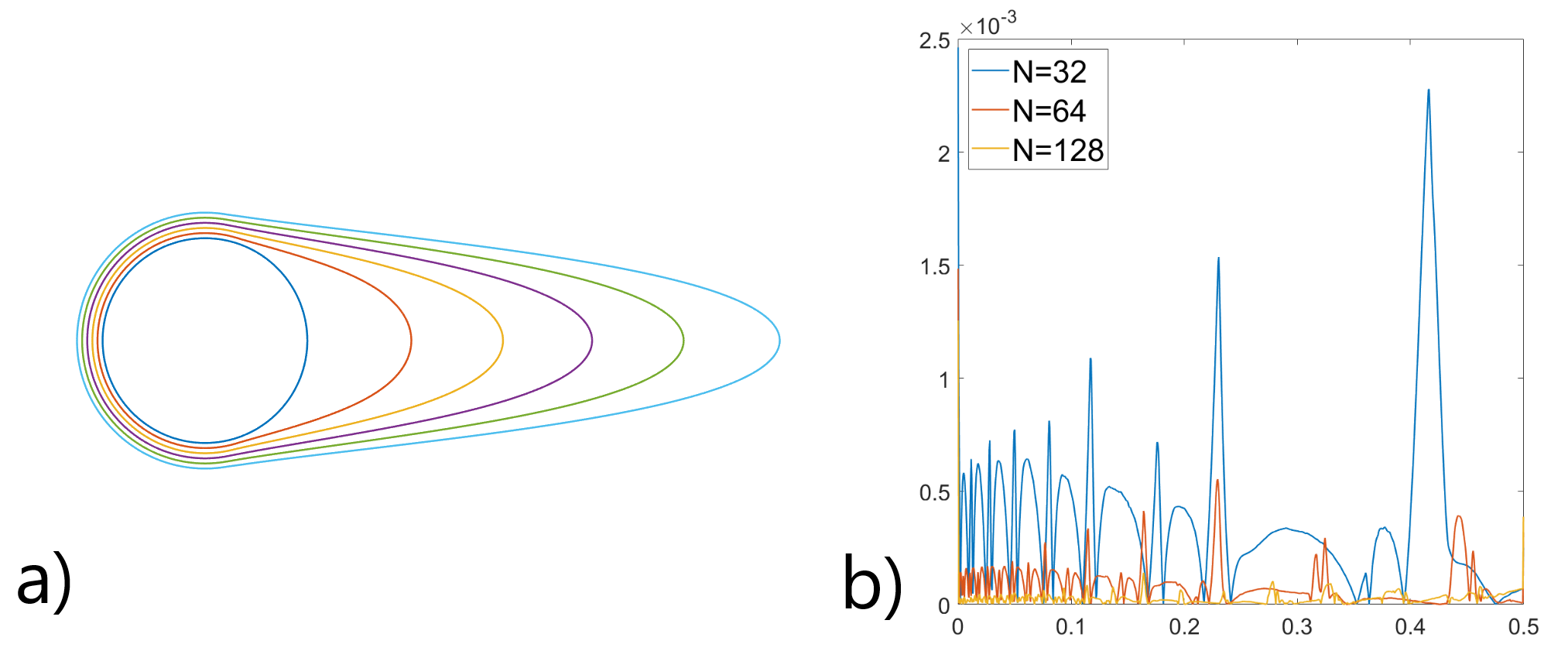}
\caption{(a) Evolution of an initially circular wildfire in a uniform ambient wind of unit strength in the positive real $z$ direction. The pyrogenic and ambient wind parameters are $\beta=10$ and $\lambda=20$ and the Laurent series truncation is $N=128$. Six isochrones at equal time increments are plotted. (b) Relative error of the RCA law of the wildfire in (a) with Laurent series truncations $N=32, 64$ and $128$.}\label{fig:RCAlaw}
\end{figure}

Consider the simplest case of an initially circular wildfire in a uniform ambient wind of strength $U_a=1$ in the real $z$ direction. The pyrogenic and ambient wind parameters are $\beta=10$ and $\lambda=20$, respectively, and the wildfire evolves for $t=[0,0.5]$. The circle is a Laurent shape and thus \eqref{Meq:LAcmap} is a good approximation for the evolving conformal map $z=f(\zeta,t)$, with a series truncation $N=128$ chosen. Figure \ref{fig:RCAlaw}a shows the wildfire evolution where six isochrones (the initial fire line and five time steps at equal time intervals) have been plotted. The evolution of the fire line is as expected: the windward side grows solely due to the (radiative) basic ROS whereas growth on the leeward side is amplified in the direction of the ambient wind and rounded into the signature parabolic shape by the pyrogenic wind \citep{sharples2022fire}. The relative error (RE) of the RCA law \eqref{eq:RCA} - see section \ref{subsec:RCA} - is computed over time and presented in figure \ref{fig:RCAlaw}b for the wildfire problem in figure \ref{fig:RCAlaw}a with Laurent series truncations of $N=32$, $64$ and $128$. As expected, this error decreases as the series truncation is increased: the average RE is of $\mathcal{O}(10^{-3})$ for $N=32$ and of $\mathcal{O}(10^{-5})$ for $N=128$. The truncation $N=128$ is thus used for all future results unless stated otherwise.

\begin{figure}[ht]
\centering
\includegraphics[width=0.92\textwidth]{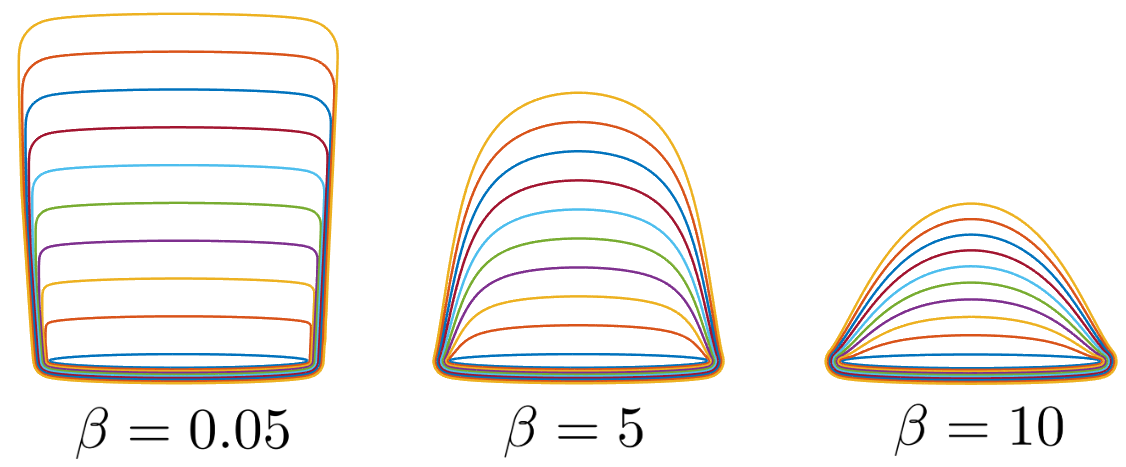}
\caption{Evaluating the effect of pyrogenic wind. Three line fires (approximated by thin ellipses) spread in the presence of a uniform, unit strength ambient wind flowing in the positive imaginary $z$ direction. The ambient and pyrogenic wind parameters are $\lambda=10$ and (a) $\beta=0.05$, (b) $\beta=5$ and (c) $\beta=10$. In each figure, ten isochrones at equal time increments are plotted.}\label{fig:pyrocomp}
\end{figure}

Following \cite{hiltonpyro} - see their figure 3 - the effect of the pyrogenic wind on the fire line propagation is investigated. Consider a straight line fire perpendicular to an ambient wind flowing in the imaginary $z$ direction. The initial line fire is approximated by a thin ellipse with conformal map \eqref{Meq:LAcmap} where $a_{-1}(0)=1$, $c_1(0)=0.9$ and $c_k(0)=0$ for all remaining $k\in[0,128]$. The fire line evolves for $t=[0,0.5]$ for a constant ambient wind parameter $\lambda=10$ and varying pyrogenic wind effects $\beta=0.05, 5$ and $10$. The isochrones of these three experiments are given in figures \ref{fig:pyrocomp}a, b and c, respectively. In agreement with the results of \cite{hiltonpyro}, it is found that a stronger pyrogenic wind effect causes an increased rounding of the leeward fire line edge into a more pronounced parabolic shape. Additionally, the fire line head does not propagate as far under stronger pyrogenic winds.

\begin{figure}[ht]
\centering
\includegraphics[width=0.9\textwidth]{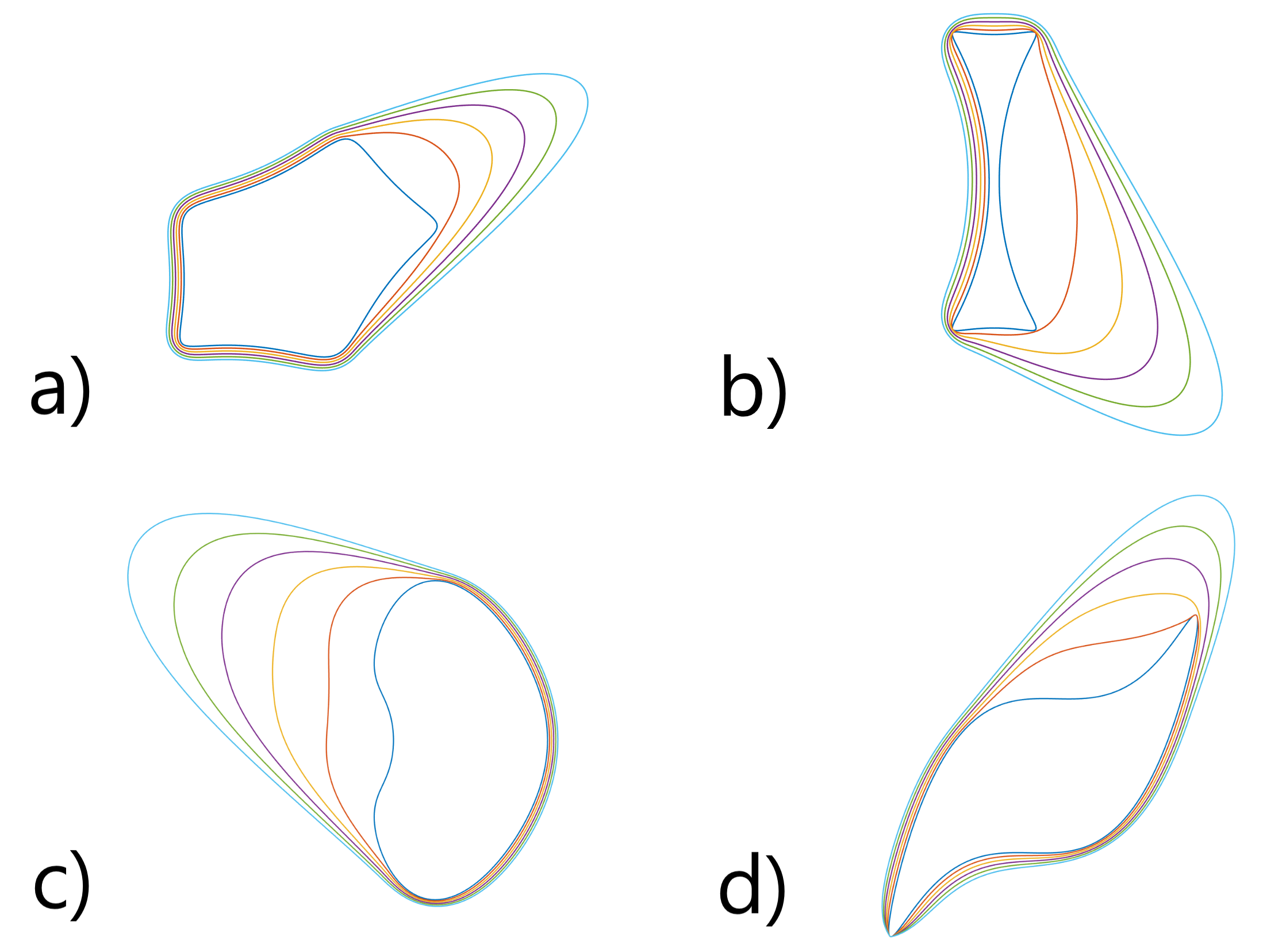}
\caption{Fire line evolution of Laurent (a,b) and non-Laurent (c,d) shapes under a constant ambient wind of strength $U=1$ at angle $\theta$ to the horizontal (real axis). (a) Irregular pentagonal fire, $\beta=10$, $\lambda=20$, $\theta=\pi/6$, $t=[0,0.25]$. (b) Hourglass fire, $\beta=18$, $\lambda=25$, $\theta=-\pi/4$, $t=[0,0.4]$. (c) Bean fire, $\beta=10$, $\lambda=20$, $\theta=5\pi/6$, $t=[0,0.2]$. (d) Blade fire, $\beta=6$, $\lambda=15$, $\theta=\pi/3$, $t=[0,0.2]$.}\label{fig:LASCshapes}
\end{figure}

Figure \ref{fig:LASCshapes} shows the evolution of a variety of initial Laurent and non-Laurent fire line shapes under the influence of different ambient wind directions and wildfire parameter values - these are given in the figure caption. Figures \ref{fig:LASCshapes}a,b are Laurent shapes: an irregular pentagon and an hourglass, respectively. These shapes and their initial Laurent coefficients come from \cite{rycroft2016asymmetric}, see their figures 5a and 7a, respectively. The remaining figures \ref{fig:LASCshapes}c,d show the evolution of initial non-Laurent fire line shapes: a bean and a blade, respectively. These come from the work \cite{gopal2019representation}, see their figure 10. Each experiment uses a series truncation $N=128$ with runtimes in the range of 5 minutes (pentagon and bean) to 35 minutes (blade) on a standard laptop. If the user desires, a smaller series truncation $N=32$ can alternatively be used to obtain quicker runtimes (on the order of seconds) while still providing a solution of good accuracy - as demonstrated in figure \ref{fig:RCAlaw}b. 

In producing the results of figure \ref{fig:LASCshapes}, a small effect on the normal velocity of the fire line proportional to the curvature of the fire line was imposed for added numerical stability. Its magnitude was of $\mathcal{O}(10^{-2})$ for Laurent shapes and $\mathcal{O}(10^{-1})$ for non-Laurent shapes; these were of sufficiently lower order compared to the basic ROS ($\mathcal{O}(1)$) and wind effects ($\mathcal{O}(10)$ such that the main features of the wildfire evolution were unchanged. The curvature effect is especially important on the windward side of the fire line where there is (generally) no pyrogenic wind contribution which also has a stabilising effect. There may be some numerical instabilities due to the crowding phenomenon which is a known challenge in numerical conformal mapping procedures, see e.g. \cite{driscoll2005algorithm}. While the $n$ points selected in the $\zeta-$plane are spaced equally around the unit disk, they may not be equally spaced in the physical plane when transformed onto the fire line. This can result in a lack of resolution at some points of the fire line, such as in deep ``troughs''. The curvature effect helps to smooth out any of these instabilities in the numerical method.

\section{Multiple spotfires model}\label{sec:modelmult}

\begin{figure}[ht]
\centering
\includegraphics[width=0.8\textwidth]{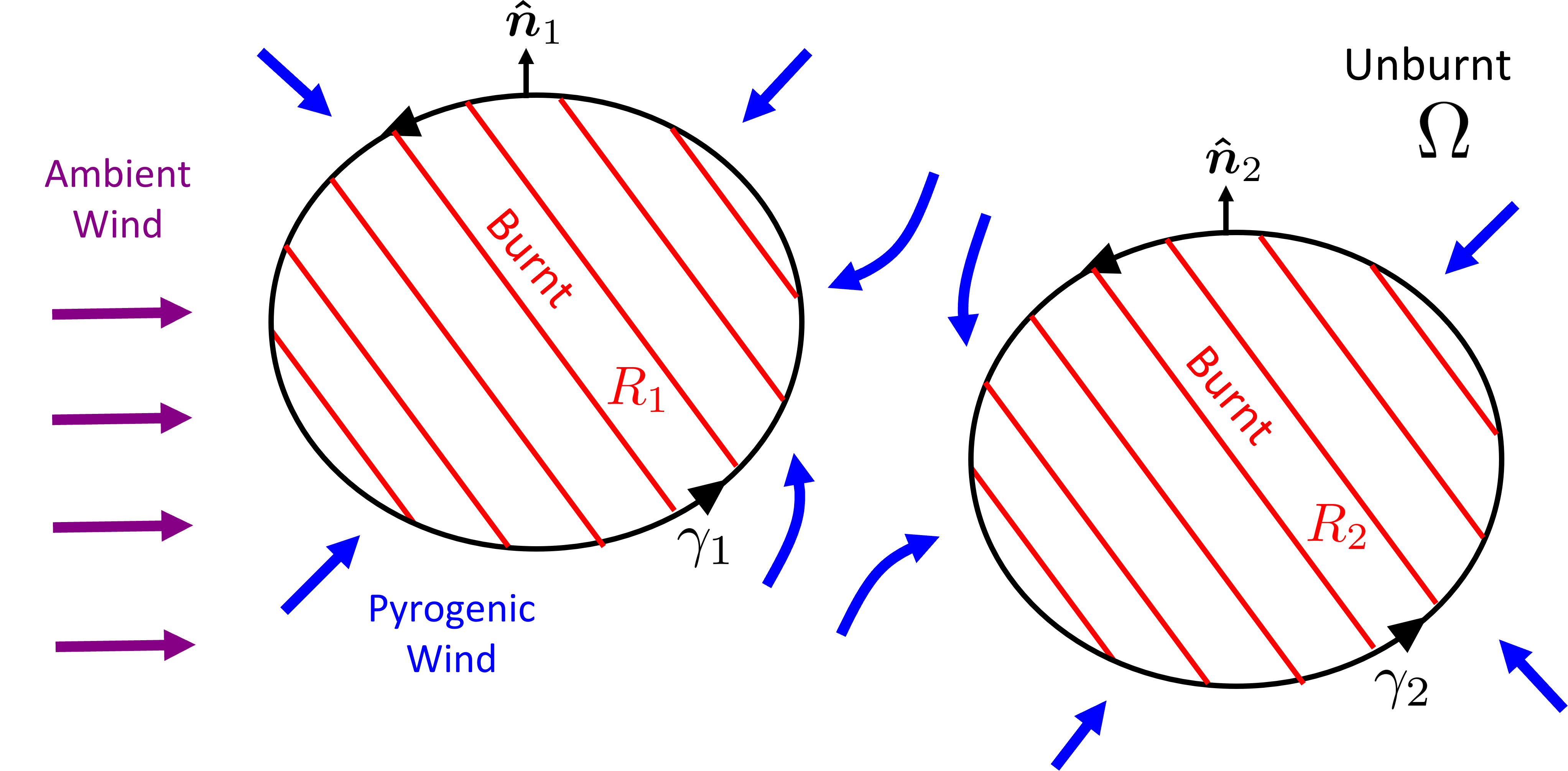}
\caption{The multiple spotfire model. The figure shows two fires from a plan view. The blue arrows represent the direction of the pyrogenic (fire) wind and the purple arrows the direction of the ambient wind.}\label{fig:multfire}
\end{figure}

Consider now the case of multiple wildfires spreading, interacting and perhaps merging with each other. These fires are again assumed to spread on a flat bed of homogeneously distributed, single-type fuel in the presence of a uniform, unidirectional constant ambient wind. These fires may have started independently or be a system of spotfires attributed to one or more main wildfires. While the stochastic generation of spotfires is a topic of  interest - see for example \cite{boychuk2009stochastic,kaur2016turbulence} and \cite{martin2016spotting} - the production of spotfires is not considered in this work. Instead, it is assumed that there are $J\geq1$ wildfires present initially and that this value $J$ will not increase over time, though it may decrease if two or more fires merge.

The total burnt region is defined as $R=R_1\cup R_2\cup\dots\cup R_J$, where $R_j$ is the burnt region of wildfire $j$ with fire line $\partial R_j = \gamma_j$, and $\Omega = \mathbb{C}\setminus R$ is the unburnt region. The velocity $[v_n]_j$ of fire line $\gamma_j$ in the direction of its outward unit normal vector $\boldsymbol{\hat{n}}_j$ is of the same form as \eqref{Meq:rcopa}
\begin{equation}
[v_n]_j=\alpha + \max[0,\;(1-\alpha) +\beta\boldsymbol{\hat{n}}_j\boldsymbol{\cdot \nabla}\phi+\lambda\boldsymbol{\hat{n}}_j\boldsymbol{\cdot \hat{u}}_a]\;\;\;\text{on }\gamma_j. \label{Meq:vn_m}
\end{equation}

The pyrogenic potential $\phi$ must be solved in the exterior region $\Omega$. The wind is assumed to be incompressible, irrotational and is the gradient of the dynamic pressure as in section \ref{sec:modelsing}, satisfying Laplace's equation \eqref{Meq:Lap} with Dirichlet boundary condition \eqref{Meq:bdry} at all $\gamma_j$. Each wildfire $j$ has a fire plume above its burned region $D_j$ and it is assumed that these plumes do not interact with each other (nor with the ambient wind as assumed in section \ref{sec:modelsing}). Therefore, each wildfire is observed in the far field as a point sink at $z=c_j$, where $c_j$ is the conformal centre of wildfire $j$. Therefore, the dimensionless system of equations governing the pyrogenic potential $\phi$ is
\begin{gather}
    \nabla^2\phi=0\;\text{ in }\Omega,\label{Meq:Lap_m}\\
    \phi=0\;\text{ on }\gamma_j,\label{Meq:bdry_m}\\
    \phi\rightarrow-\log{|\Pi_{j=1}^J(z-c_j)|}=-\log{|h(z)|}\;\text{ as } r\rightarrow\infty. \label{Meq:far_m}
\end{gather}
Equations \eqref{Meq:vn_m} - \eqref{Meq:far_m} have been non-dimensionalised using the same scalings \eqref{eq:scalings} as in section \ref{sec:modelsing}. The characteristic length scale $R_0$ is now the conformal radius of the largest initial wildfire and the velocity scale is still $v_0$: the basic ROS of the fuel. Further, it is assumed that each wildfire generates a fire plume of constant strength $Q$ which is used in scaling the potential $\phi$. This is  a simplification, and an alternative scenario not explored here is that smaller fires may have weaker plumes and thus a smaller effective $Q$. The starting fires in this work are all of comparable size.

\section{AAA-least squares numerical method}\label{sec:aaa}
The single wildfire numerical method from section \ref{sec:cmap} cannot be used in the multiple spotfires scenario as the conformal map from the multiply connected domain to some canonical domain (which is no longer the simple unit disk) cannot be written as a simple Laurent series for all time. Therefore, an alternative method is needed. This would involve explicit time stepping which adds to the computation cost, thus quick methods are sought at each time step to make this procedure more appealing for operational use. There are two processes to consider in constructing this numerical method: how the pyrogenic potential $\phi$ in the exterior region $\Omega$ is calculated at each time step and; how  two or more overlapping wildfires are detected and merged into one new wildfire.

\subsection{Calculating the pyrogenic potential}
In the last five years, the AAA-LS algorithm has been created and developed for solving two-dimensional Laplace problems, see the recent review article by \cite{nakatsukasa2023first}. The algorithm is based upon the method of fundamental solutions (or the ``charge simulation'' method, see e.g. \cite{amano1994charge}) and combines the AAA algorithm for finding a rational approximation to some boundary data \citep{costa2020solving} with a least squares (LS) fitting procedure, such as in the Lightning Laplace solver \citep{gopal2019solving}. This method is designed to handle non-smooth domain boundaries with corner and cusp singularities and can be extended to the domain exterior as well as to multiply connected domains \citep{trefethen2020numerical}; all of these extensions are of interest in this work.

The harmonic pyrogenic potential $\phi$ in the unburnt region $\Omega$ can be approximated by $\phi=-\log{|h(z)|}+\text{Re}[F(z)]$, where $F$ is some complex, analytic function in $\Omega$ \citep{trefethen2020numerical}. The function $F$ can be approximated as
\begin{equation}\label{eq:Fmult}
    F(z) \approx \sum_{j=1}^J\bigg[D_j\log{\bigg{(}\frac{z-c_j}{z-c_{j'}}\bigg{)}}+\sum_{n=1}^N A_{jn}(z-c_j)^{-n}+\sum_{p_{jk}\in D} \frac{B_{jk}}{z-p_{jk}}\bigg],
\end{equation}
where $j'=j\pmod{J}+1$ and $A_{jn}$, $B_{jk}$ are unknown complex coefficients. This approximation consists of three terms. First, a sum of logarithmic terms with unknown real coefficients $D_j$, where $\sum_{j=1}^J D_j = 0$ is implicitly imposed such that there is non-singular behaviour as $z\rightarrow\infty$ \citep{trefethen2018series,costa2023aaa}. Second, a smooth (Runge) polynomial term truncated at $N$; the value $N=20$ is used in this work. Third, a singular (Newman) part consisting of simple poles $p_{jk}$ clustered exponentially near corner and cusp singularities in the burnt region i.e. the unphysical region of the problem. While these poles could be placed manually, the AAA algorithm is instead used to generate suitable poles based on the far field function $\log{|h(z)|}$. Any poles of tiny residue $<\mathcal{O}(10^{-8})$ or sufficiently close to the fire line where $\phi=0$ (at distances less than $\mathcal{O}(10^{-2})$) are then manually excluded to avoid cases of spurious poles or ``Froissart doublets'' \citep{nakatsukasa2018aaa}.

The vector of unknown coefficients $c = [A_{jn}; B_{jk}; D_j]$ is found using an LS algorithm applied to the boundary data $z_b$, utilising the boundary condition $\text{Re}[F(z_b)]=\log{|h(z_b)|}=H(z_b)$ on $\gamma_j$. Constructing the Vandermonde matrix A of basis vectors (with Arnoldi orthogonalisation included for improved stability - see \cite{brubeck2021vandermonde}) gives the expression $Ac=H$ and thus $c$ can be found using the backslash operator $c=H\backslash A$. Finally, using the relation $\nabla\phi=\overline{F'(z)}$ from \cite{trefethen2018series}, where $\nabla$ is the complex representation of the corresponding vector, it follows that
\begin{equation}\label{eq:pyrocont}
\boldsymbol{\hat{n}}_j\boldsymbol{\cdot\nabla}\phi = \text{Re}[\hat{n}_j\overline{\nabla}\phi]=\text{Re}[\hat{n}_j(-h'(z)/h(z)+F'(z))],
\end{equation}
which is the pyrogenic wind effect given in \eqref{Meq:vn_m}. Thus the normal velocity $[v_n]_j$ can be computed for each fire line at a given time step $t$: the values $\alpha,\beta,\lambda$ and $\boldsymbol{\hat{u}}_a$ are known constants and the normal vector $\boldsymbol{\hat{n}}_j$ of each fire line can be computed from the boundary data $z_b$. 

The AAA-LS algorithm runs at each time step, with the fire line then propagated through time using standard Runge-Kutta (RK) methods, see e.g. \cite{butcher1996history}. These methods have an accuracy of $\mathcal{O}(\Delta t^n)$ where $n$ is the order of the RK algorithm used, with the AAA-LS method used $n$ times at each time step. In this work, fourth-order Runge-Kutta (RK4) is generally used with an $\mathcal{O}(10^{-2})$ time step. However, the first-order Euler's method (RK1) may be used for simple scenarios (eg circular wildfires) where a similar level of accuracy can be achieved using the same order of time step. The AAA-LS algorithm with RK time stepping offers the speed desired, with an RK1 time step running in around two seconds and RK4 in 8 seconds on a standard laptop; this is comparable with previous runtimes of the AAA-LS algorithm stated in e.g. \cite{trefethen2020numerical,costa2023aaa}. Further, the AAA-LS method offers root-exponential convergence with respect to the number of poles generated \citep{gopal2019solving}. With $\mathcal{O}(100)$ poles produced on each run of the AAA algorithm, solutions provided are also of suitable accuracy.

\subsection{Detecting and merging two overlapping fire lines}\label{subsec:mergalg}
The merging (and splitting) of two free boundaries is a common problem in vortex dynamics (see e.g. \cite{xue2017new}) where the boundaries represent the interface between rotational and irrotational flow. The numerical method of contour surgery pioneered by \cite{dritschel1988contour} has been successful in enabling complex vortex interaction to be studied. Typically, the numerical problem involves calculating the minimum distance between the two free boundaries and, when this distance falls below some tolerance level, drawing two additional segments to connect the two contours into a single, new closed curve. The MATLAB function $union$ used in this work offers a simple approach, with the function correctly ordering and orienting the boundary points of the new merged fire line in the anticlockwise direction. However, the $union$ function  will only merge the two fire lines if the curves are overlapping and are not just sufficiently close to each other. This is of no consequence to the AAA-LS algorithm which can easily calculate the pyrogenic potential $\phi$, even when the fire lines are close together - the potential is very small in the gap region and hence is effectively zero there. 

As soon as the fire lines overlap, they are then merged into a single contour. However, the RK4 method will only merge the fire lines (if applicable) at the end of the time stepping procedure. Thus this method will fail if the fire lines overlap during the time step as the AAA-LS algorithm is not designed to handle overlapping contours. Therefore, if a merge occurs during a RK4 step, the code will reset to the beginning of the time step and perform a simple RK1 time step instead. To allow the solution to achieve the same level of accuracy throughout, the RK1 time step will be sufficiently smaller than the one used for the higher order RK method; in this work, an ``emergency'' RK1 step size of $\mathcal{O}(10^{-4})$ is typically used. This is another reason why RK1 time stepping may be used instead of RK4: the isochrones of RK4 may not be at equal time intervals between each other if any emergency RK1 steps were used. Therefore RK1 is useful for ensuring that all isochrones are plotted at equal time intervals and thus for comparing isochrones at specific times - this is particularly important in section \ref{sec:hcomp} when comparing the model outputs with experimental data.

Some additional procedures are imposed at each time step over the entire numerical process for added numerical stability. The function \textit{self-intersect} created by \cite{canos2024selfintersect} identifies any segments of the fire line which have self-intersected between time steps and removes them from the list of boundary points such that the fire line remains a simple, singly connected curve. Self-intersection is common in the case of extreme junction fires ie V-shaped fires (see e.g. \cite{viegas2012study,ribeiro2023slope}) for example those resulting from a fire merger. If the \textit{self-intersect} procedure occurs during an RK4 time step, the time step is aborted and a smaller, emergency RK1 time step is instead performed, as is also employed in the merging procedure. The MATLAB function \textit{smoothdata} smooths any noise in the fire line data without affecting the overall shape of the fire line. Such provisions are necessary for two reasons. First, to allow the user to use a larger time step and still obtain a numerically stable solution; this means the code is computationally quicker and cheaper which is vital for operational use. Second, while the pyrogenic wind is a stabilising effect, the entropy condition in \eqref{Meq:vn_m} means that there is no stabilising behaviour on the leeward side of the fire or (often) in the valley between fires. In the single wildfire case, a small magnitude stabilising curvature effect was used when necessary to impose the stability in these regions. In the multiply connected case, where the explicit computation of curvature could also create numerical errors, the \textit{selfintersect} and \textit{smoothdata} functions are used instead. Finally, the function \textit{interparc} created by \cite{derrico2024interparc} interpolates the polygon at each time step to maintain the same resolution of points on the fire line as the wildfire grows.

\section{Two and three wildfire results}\label{sec:resultsmult}
\begin{figure}[ht]
\centering
\includegraphics[width=0.99\textwidth]{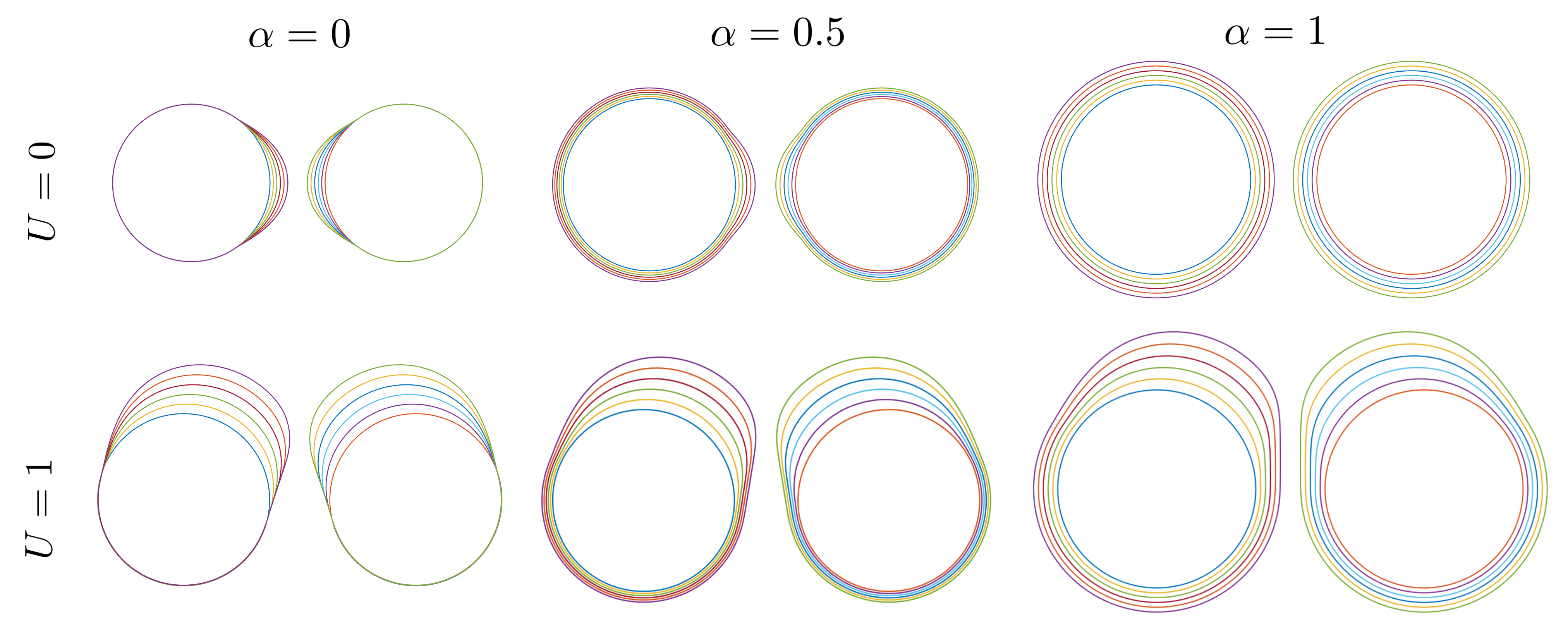}
\caption{Six independent subfigures showing the growth of two circular wildfires with (bottom row) and without (top row) a unit strength ambient wind in the positive imaginary $z$ direction for the values $\alpha=0$ (left), $\alpha=0.5$ (centre) and $\alpha=1$ (right). Each fire grows with $\beta=1.5$ and $\lambda=3$ for $t=[0,0.25]$, with isochrones plotted at $t=0.05$ increments.}\label{fig:twocircles}
\end{figure}

Figure \ref{fig:twocircles} gives the simplest multifire example of two growing circular wildfires, interacting with each other before merging occurs. Six independent scenarios are presented. The bottom and top rows of figure \ref{fig:twocircles} show the growth of the fires with and without a unit strength vertical ambient wind present, respectively. The three columns from left to right show the effect of increasing the value of $\alpha$ - the ratio of radiative to convective basic ROS - from $0$ (fully convective) to $0.5$ (the representative value taken throughout this work) to $1$ (fully radiative). In each figure, the values $\beta=1.5$ and $\lambda=3$ are selected for illustrative purposes. Further, the fires evolve in the time $[0, 0.25]$ with isochrones plotted at intervals $t=0.05$, with RK1 timestepping used to allow more direct comparison between the fire scenarios at specific time increments. Each result was created in under four minutes of runtime.

The subfigures of figure \ref{fig:twocircles} show distinctive behaviour. Considering $U=0$, the example $\alpha=0$ shows fire growth only in the region between the two wildfires - the incoming pyrogenic wind is sufficiently strong enough on the remaining sections of each fire line to stop fire spread entirely, so that the most pronounced effect is the wildfires growing towards each other. The other extreme $\alpha=1$ shows the two fires growing entirely independently of each other, with the pyrogenic wind effect completely ignored at all points of the fire line due to the entropy condition. However, this is inconsistent with the observations and conclusions drawn in \cite{hiltonpyro} which state that two wildfires will interact and grow towards each other (even in the absence of ambient wind). Careful consideration has consequently been taken in this work to separate radiative and convective basic ROS effects in the wildfire model as neglecting either effect results in two (unphysical) extreme cases. The example where $\alpha=0.5$ gives a good balance between these two extremes: the fires grow towards each other but convective effects are not strong enough to entirely stop fire spread on non-facing segments of the fire line. The examples with ambient wind $U=1$ in the imaginary $z$ direction show much the same effects, now with the added rounding of the fire head on the leeward side. Additionally, there is evidence even in the $\alpha=1$ case that the wildfires grow towards each other. However, this is only seen at the fire head; there is no apparent interaction (as mathematically expected) in the region between the wildfires. Once more, this is inconsistent with theoretical work and experimental observations - see section \ref{sec:hcomp}.

\begin{figure}[ht]
\centering
\includegraphics[width=0.8\textwidth]{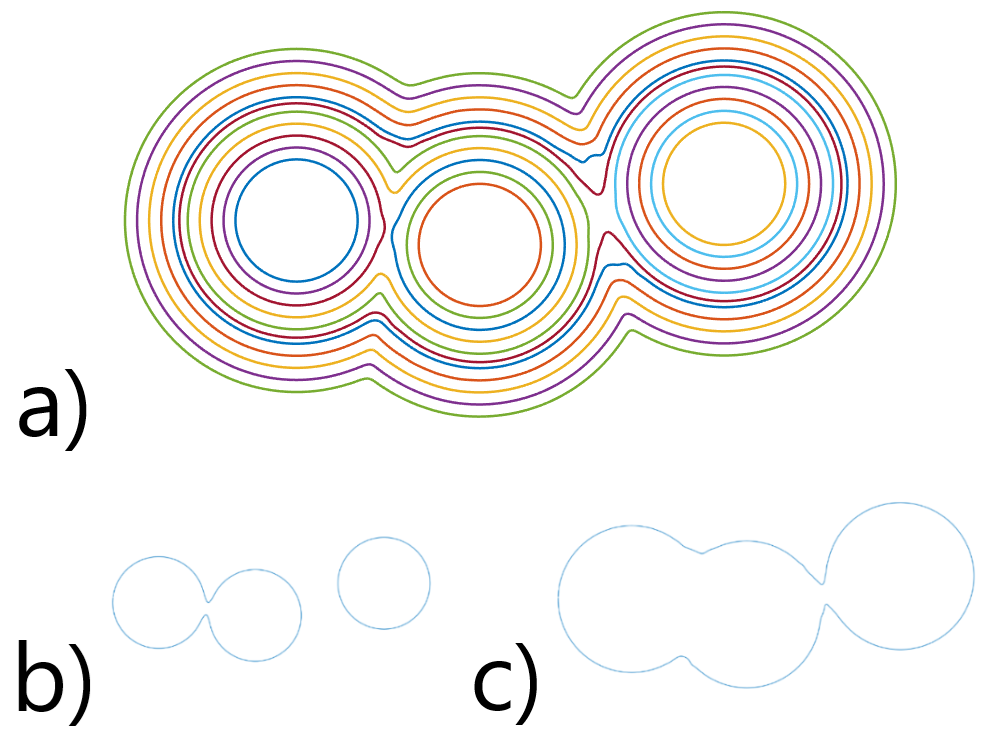}
\caption{The spread and merger of three circular wildfires with $\beta=10$ in the absence of ambient wind. Of the 100 isochrones calculated, 20 are plotted in (a), with the contours of the first (b) and second (c) mergers identified.}\label{fig:threecircles}
\end{figure}

Figure \ref{fig:threecircles} shows the growth and eventual merger of three circular wildfires in the absence of ambient wind. The parameter $\beta=10$ is used and RK4 timestepping is employed, with time steps $t=0.02$ for each RK4 step and $t=5\times 10^{-4}$ for any emergency RK1 time steps. A total of 100 steps are performed with 20 of these steps plotted in figure \ref{fig:threecircles}a; note that these isochrones are now not necessarily at equal time intervals due to the use of RK1 at times of merger and numerical instability. The first and second merger steps are displayed in figures \ref{fig:threecircles}b and \ref{fig:threecircles}c, respectively, and all results were produced in $28$ minutes of runtime. The resulting wildfire growth is as expected: the fires grow more strongly towards each other as they approach and the concave regions of the fire line following a merger are quickly smoothed out, with both of these effects attributed to the pyrogenic wind. Further, the overall fire evolution looks smooth, with any numerical instabilities successfully smoothed out while not changing the main features of the fire spread. This is observed between the second and third isochrones after the second merger: the temporary instabilities in the fire junction are smoothed away.

\section{Firebreaks}\label{sec:fbreak}
The numerical method of this work is also able to model the effects of firebreaks, or more generally segments of the fuel bed which are unable to burn, for example roads, rivers, lakes or areas of previous controlled burning. It is assumed that the firebreaks are sufficiently flat so that both ambient and pyrogenic winds are unaffected by their presence. Consider the generalised multifire scenario of $J$ wildfires with burnt regions $R_j$ and fire lines $\gamma_j$, respectively, with the unburnt region defined as $\Omega=\mathbb{C}\setminus [R_1\cup R_2\cup\dots\cup R_J]$. Initially and therefore for all time, there is a (perhaps disconnected) region $I$ consisting of incombustible fuel which is a subset of the unburnt region $\Omega$. A very simple addition is made to the normal velocity equation \eqref{Meq:vn_m} on each fire line $\gamma_j$ as follows
\begin{equation}
[v_n]_j=\tau(z)\Big[\alpha + \max[0,\;(1-\alpha) +\beta\boldsymbol{\hat{n}}_j\boldsymbol{\cdot \nabla}\phi+\lambda\boldsymbol{\hat{n}}_j\boldsymbol{\cdot \hat{u}}_a]\Big],\;\;\;\;\;\tau(z) = 
    \begin{cases}
        0,\;\;\; z\in I\\
        1,\;\;\; z\notin I.
    \end{cases} \label{Meq:vn_tau}
\end{equation}
Although not considered here, inhomogeneous fuel beds could also be described using a similar equation, where sections of the fuel bed speed up or slow down wildfire spread rather than stop it altogether. In such a scenario, consideration must be made whether this would affect the entire normal velocity or simply the basic ROS contributions.  The modified normal velocity equation \eqref{Meq:vn_tau} is implemented into the multiple wildfires model and various scenarios considered - there is zero ambient wind in each scenario but pyrogenic wind is still present.

\begin{figure}[ht]
\centering
\includegraphics[width=0.65\textwidth]{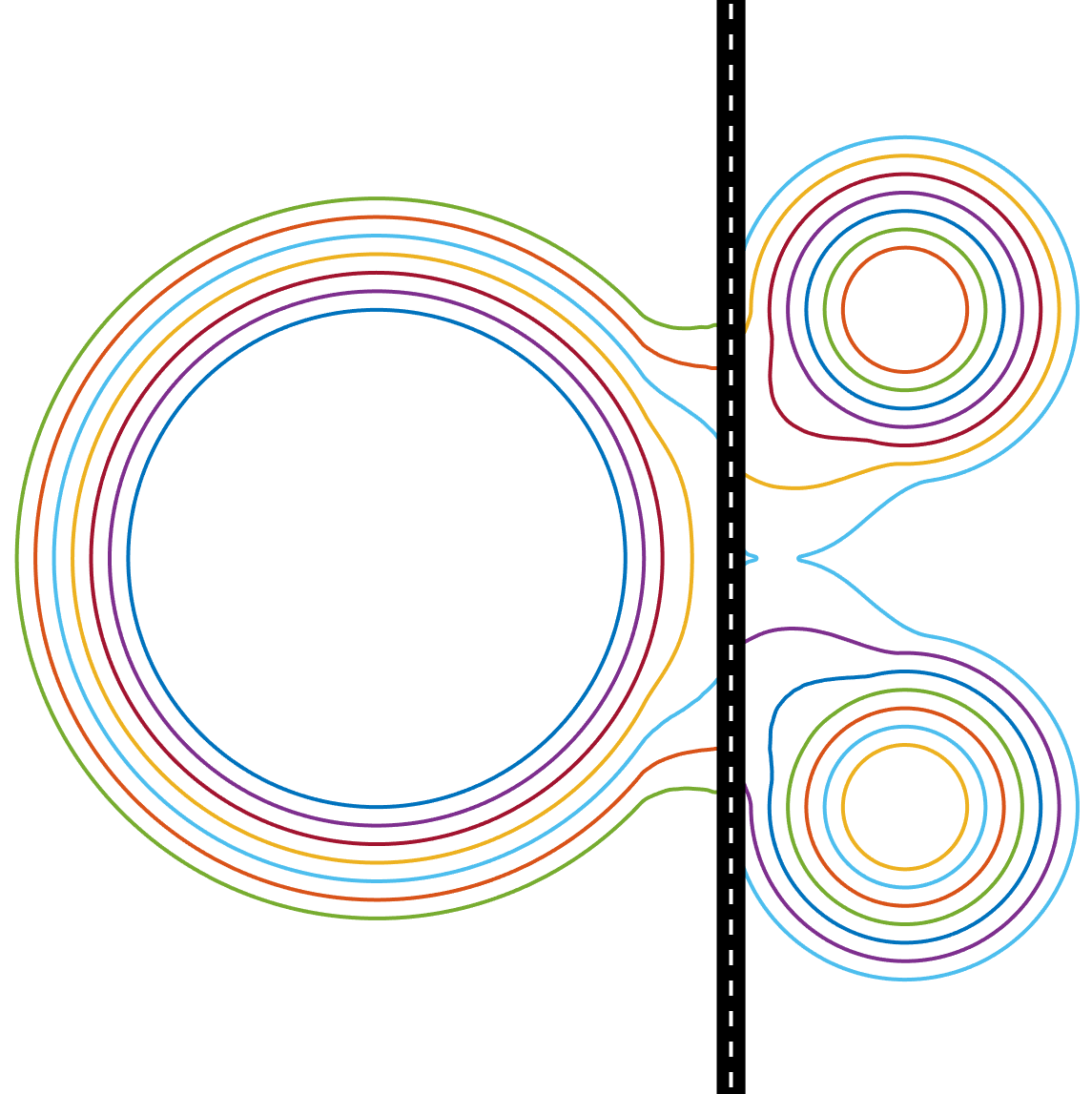}
\caption{Three fires on either side of a straight road - the infinite strip. Ten isochrones are plotted showing the fire spread at equal time increments.}\label{fig:tcr}
\end{figure}

Figure \ref{fig:tcr} shows an example involving three initially circular fires on either side of a straight road which is modelled by an infinite strip parallel to the imaginary direction. The motivation for this example is that the two fires on the right of the road are smaller spotfires produced from the main wildfire on the left. As stated in section \ref{sec:resultsmult}, the generation of these spotfires is not discussed in this work. Ten of the 73 isochrones calculated are plotted, where RK1 with a time step of $t=0.05$ is used. The parameters $\beta=10$ and $\alpha=0.25$ are used for illustrative purposes and the code runs in 770 seconds. None of the three fires penetrate into the road and thus the numerical method successfully accounts for the firebreak. Further, the three fires continue to grow towards each other due to the pyrogenic wind - note the lower value of $\alpha=0.25$ is taken here to emphasise this effect. The two spot fires grow towards the main fire and each other and eventually merge.

\begin{figure}[ht]
\centering
\includegraphics[width=0.95\textwidth]{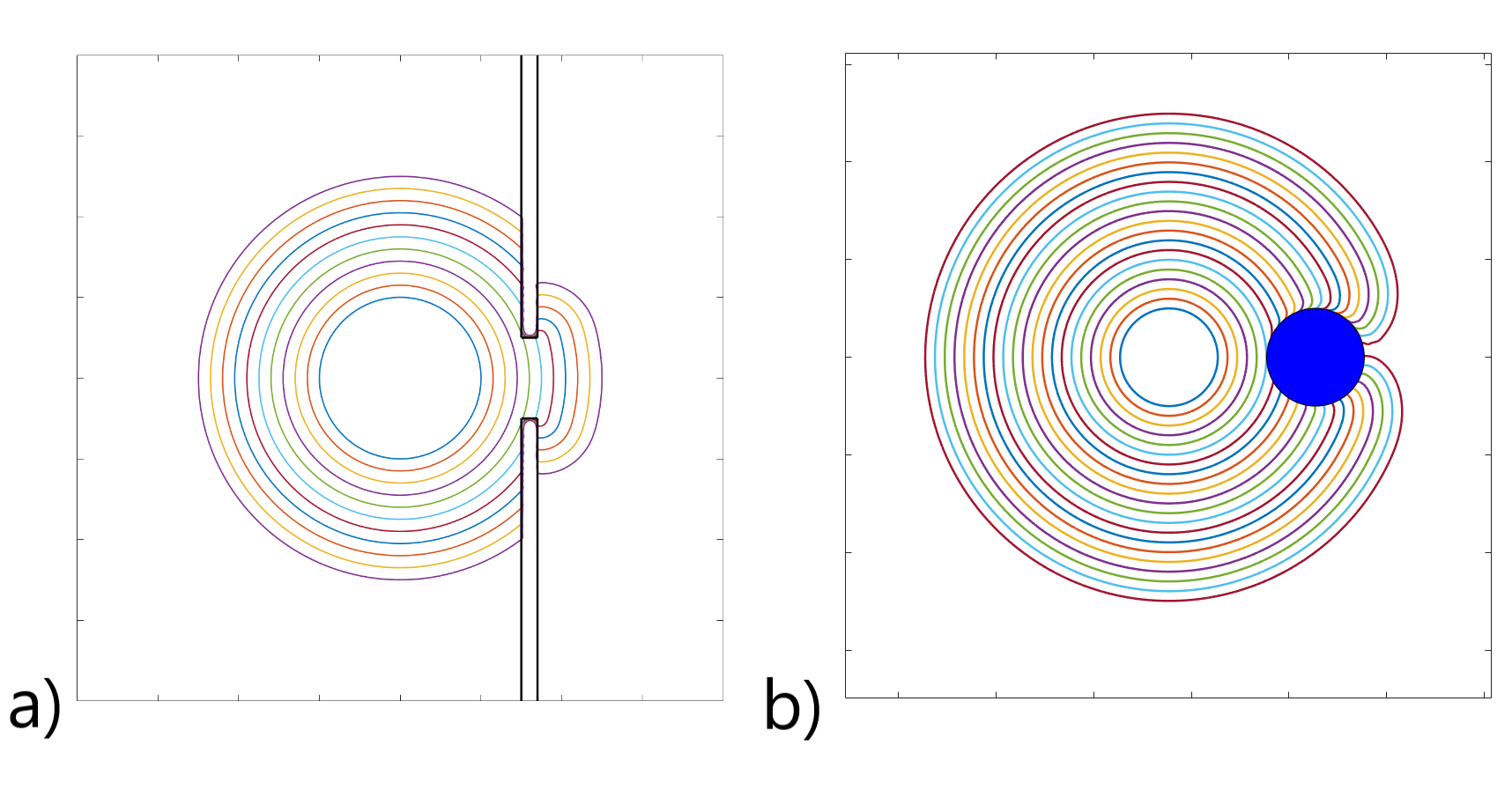}
\caption{Evolution of an initially circular wildfire in the presence of various fire breaks. Isochrones of the wildfire evolution (coloured lines) are plotted in each figure. (a) A gap in the firebreak - the infinite strip between the black lines. (b) A circular lake - the filled, blue circle.}\label{fig:gaplake}
\end{figure}

Two further firebreak examples involving only a single, initially circular wildfire are given in figure \ref{fig:gaplake}. The first in figure \ref{fig:gaplake}a shows the previous infinite strip scenario with an opening in the middle; this could represent a gap between two purpose-built firebreaks designed to slow the fire spread rather than stop it altogether. The parameter $\beta=20$ is taken and RK1 is used with time step $t=0.1$ for 30 steps, running in a time of 122 seconds. The wildfire successfully penetrates through the gap in the firebreak, yet the radius of the wildfire spread on the right side is largely reduced from that on the left, as expected. Figure \ref{fig:gaplake}b shows the spread of the wildfire with $\beta=10$ around a circular lake - the filled, blue circle of radius $0.5$. Here, RK4 is used with time step $t=0.1$ for 40 steps, running in a time of $570$ seconds. The wildfire splits into two arms which grow around the lake and towards the other arm, with the simulation stopped before the two arms merge. The multifire model in this work cannot account for this type of merger as it would result in two disconnected ``exterior'' domains, yet such an extension to the AAA-LS algorithm is possible; this is discussed in section \ref{sec:conc}.

\section{Comparison with results from \cite{hiltonpyro}}\label{sec:hcomp}
The single wildfire and multiple spotfires models developed in this work are inspired by the model of \cite{hiltonpyro}, with some notable differences as discussed in section \ref{sec:intro}. Thus it is natural to test the accuracy and robustness of the models of this work by reproducing some of the results from \cite{hiltonpyro}. In particular, their figures 5, 7 and 8 are considered which show the evolution of a line fire and a connected and unconnected junction fire\footnote{This is a V-shaped fire composed of two line fires at an angle $\theta$ to each other; the fires considered here have $\theta=\pi/2$.}, respectively, all in the presence of an ambient wind flowing in the positive imaginary $z$ direction. The results of the pyrogenic potential model of \cite{hiltonpyro} are then compared with experimental fire data performed in the CSIRO Pyrotron - see \cite{sullivan2019investigation} for the full details of each experiment. The figures and additional data from these experiments, in particular further time steps of figure 8, were kindly provided by the authors of \cite{hiltonpyro} and \cite{sullivan2019investigation} for use in this work.

All three experiments use the same parameters and so the same (equivalent) parameters are to be found and used here. As in \cite{hiltonpyro}, the basic ROS is $v_0 = 5\times10^{-4}$ms$^{-1}$ and the ambient wind is of strength $U_a=1$ms$^{-1}$ flowing in the positive imaginary $z$ direction. As in section \ref{sec:modelsing}, it is taken that $\alpha=0.5$ and $\lambda=2\beta=\mathcal{O}(10)$, with the value $\lambda=20$ found to best match the results. The experiments in figures 5 and 7 of \cite{hiltonpyro} run for $t=15$ seconds whereas figure 8 runs for $t=40$ seconds. It was found that the value $A=0.24$ gives the best fit to all results, recalling the dimensionless number $A$ from section \ref{sec:modelsing} which scales the time, see equation \eqref{eq:scalings}. Each line fire has initial length $0.8$m, yet recall from section \ref{subsec:nondim} that the $R_0$ scaling used here is the conformal radius (of the largest initial fire). The SC Toolbox was used to calculate the initial conformal radii for each experiment, noting that the fire line at dimensional time $t=5$ seconds is taken to be the initial fire line here. The scales of the figures are also taken into account when calculating $R_0$ - figure 5 is displayed on a $1$m$\times1$m grid and figures 7 and 8 on a $1.4$m$\times1$m grid. The initial conformal radii for figures 5, 7 and 8 of \cite{hiltonpyro} are $R_0=0.25$m$, 0.38$m and $0.24$m, respectively. 

The dimensionless final time $t_{max}$ for each experiment can then be calculated using the scaling in equation \eqref{eq:scalings}. For the single fire scenarios of figure 5 and 7, a Laurent series truncation of $N=128$ is used and an $\mathcal{O}(10^{-2})$ curvature effect is included for added stability. For the two fire scenario of figure 8, RK1 is used with a time step of $t=4.4\times10^{-3}$ for 800 steps such that isochrones at specific times can be compared with the figures of \cite{hiltonpyro} - see section \ref{subsec:mergalg} for more details. As these results are produced for comparative purposes, a longer overall runtime (which is a consequence of using RK1 with a smaller time step) is acceptable. Finally, all relevant parameters are scaled by a factor $10^{-1}$, e.g. $\lambda=2$ rather than $20$, to allow for greater numerical stability. It should again be noted that the same parameters are used for all three experiments here and care has been taken to ensure they match as well as possible with the parameters used in \cite{hiltonpyro}.

\begin{figure}[ht]
\centering
\includegraphics[width=0.8\textwidth]{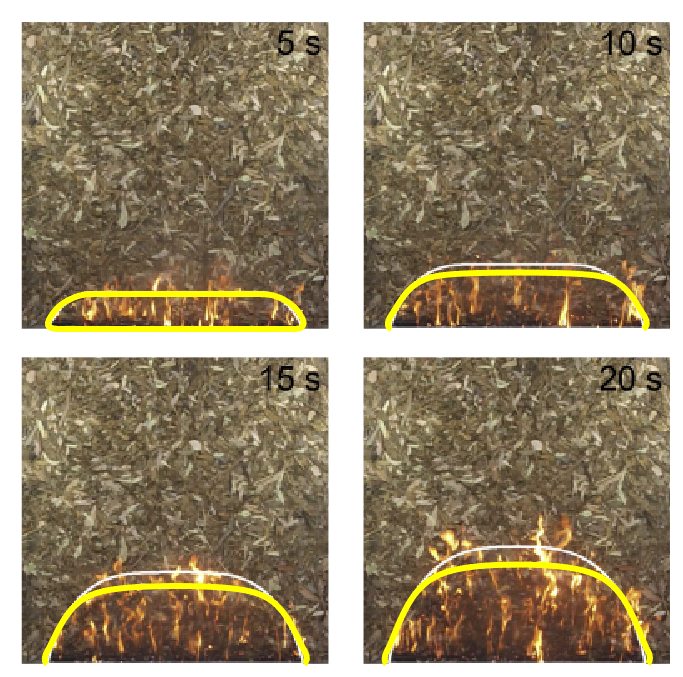}
\caption{Comparison with figure 5 of \cite{hiltonpyro}.}\label{fig:hilton18fig5}
\end{figure}

\begin{figure}[ht]
\centering
\includegraphics[width=0.88\textwidth]{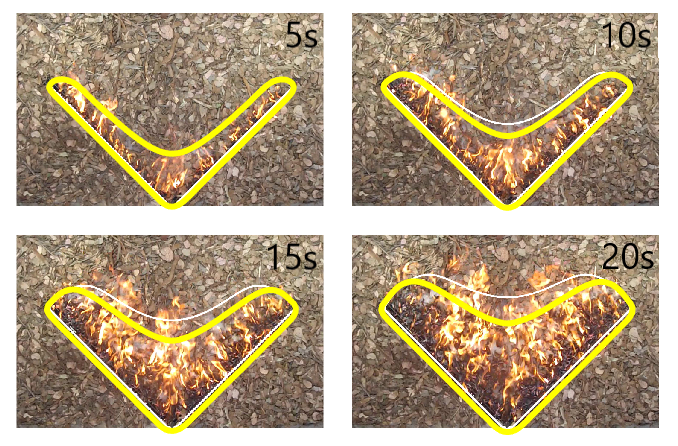}
\caption{Comparison with figure 7 of \cite{hiltonpyro}.}\label{fig:hilton18fig7}
\end{figure}

\begin{figure}[ht]
\centering
\includegraphics[width=0.95\textwidth]{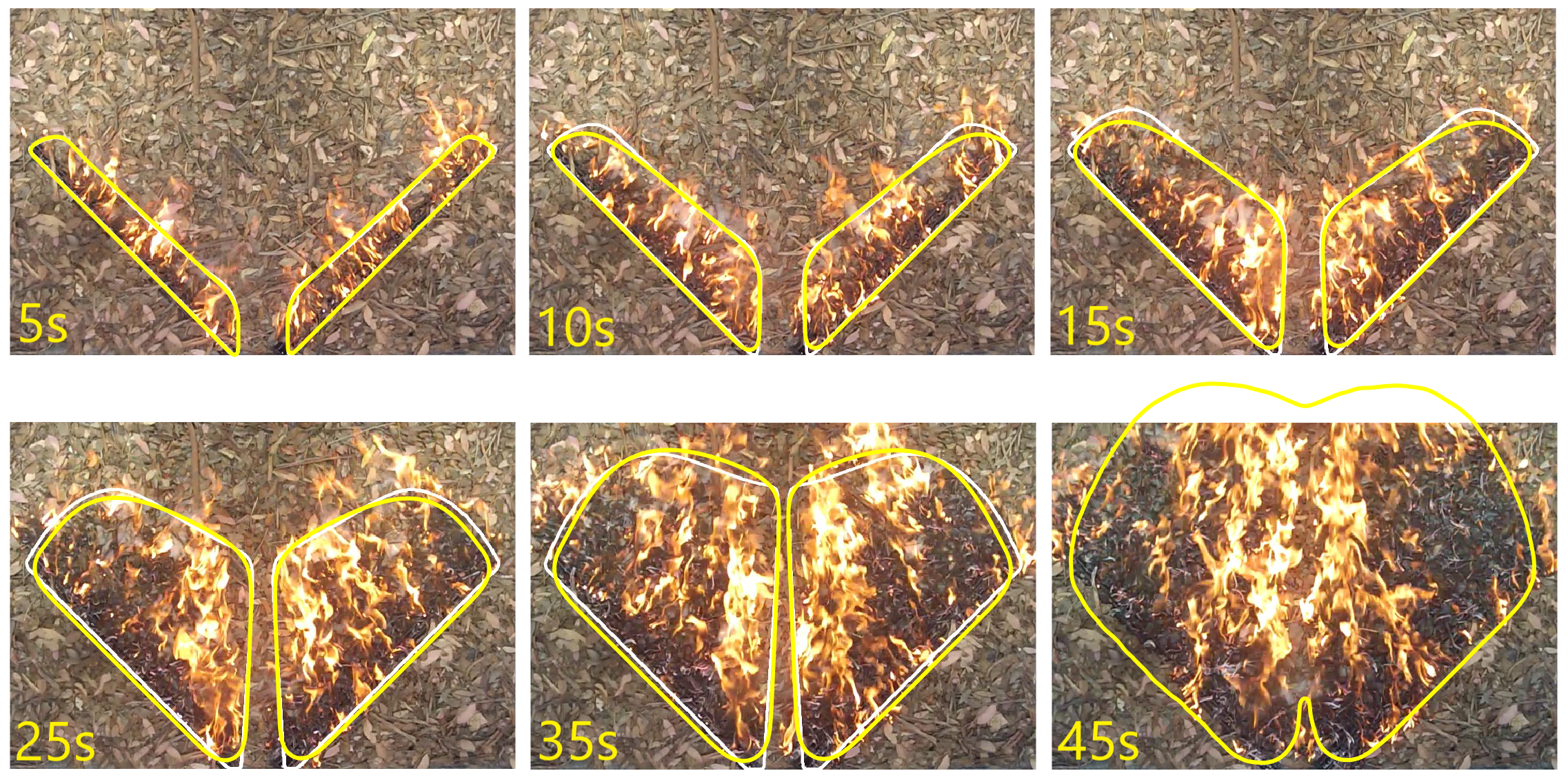}
\caption{Comparison with figure 8 of \cite{hiltonpyro}.}\label{fig:hilton18fig8}
\end{figure}

Figures \ref{fig:hilton18fig5}, \ref{fig:hilton18fig7} and \ref{fig:hilton18fig8} are analogues of figures 5, 7 and 8 from \cite{hiltonpyro}, respectively, and superimpose the results from the code of this work (yellow isochrones), the results from the pyrogenic potential model of \cite{hiltonpyro} (white isochrones) and the experimental data from \cite{sullivan2019investigation} (background image) all onto the same plot. Figures \ref{fig:hilton18fig5} and \ref{fig:hilton18fig7} show the development of a single  line fire and a single connected junction fire, respectively, both of which are perpendicular to the ambient wind. Each scenario had a runtime of around $27$ minutes, though there was little difference to the $N=32$ results which ran in $15$ seconds. The model of this work successfully captures the main features of the fire development in both experiments: the development of the fire line head into a parabolic shape in the windward direction in figure \ref{fig:hilton18fig5} and the closing up of the junction fire in figure \ref{fig:hilton18fig7}. Further, the fire line shapes qualitatively agree with those from the model of \cite{hiltonpyro}, though quantitatively the fire lines differ more noticeable at larger times. However, there is still good agreement between the model of this work and the experiment; the model still appears to enclose the burnt region of the fire at all times. Neither the model of this work nor the \cite{hiltonpyro} pyrogenic potential model can account for the 3D projection of flames forward of the fire and so, even though the fire line successfully encloses the burnt region, there is still evidence of flames extending beyond the fire line.

Figure \ref{fig:hilton18fig8} shows the development of an initially  disconnected junction fire - two line fires angled towards each other - which had a runtime of around 50 minutes. There is once again good agreement between the model of this work, the experimental fire and the model of \cite{hiltonpyro} over all time. There is some evidence of ``over-rounding'' of the fire line: the eastern and western flanks are more curved than in \cite{hiltonpyro} and the southern-most points of the wildfire round out also. This is a consequence of the smoothing processes, namely the function \textit{smoothdata} in section \ref{sec:aaa}, applied during the multifire code to help eliminate numerical instabilities. These small errors are acceptable given the aim of this work to produce an operational model capable of capturing the main features of the wildfire spread. Further, the final fire line at time $t=45$ seconds reproduces the burnt region well, noting that the two line fires have successfully merged by this time. Unfortunately the image is cut off such that the head of the fire is not visible in the photo, nor is there a comparison with the work of \cite{hiltonpyro} at this final time step. However, it is concluded that the model of this work shows good agreement with the experimental fire for all time.

\section{Discussion}\label{sec:conc}
Wildfire spread, spotfire merger and the effect of firebreaks have been modelled in this work using a simple, two-dimensional model. The wildfire spread itself is dependent on three main effects: the basic rate of spread divided explicitly into its radiative and convective components; the ambient wind and; the pyrogenic wind - a self induced wind caused by the updraft created by the wildfire over its burnt region. While in this model the ambient wind and basic ROS effects are purely kinematical and prescribed by given constants, the pyrogenic wind is determined by the solution of Laplace's equation in the wildfire exterior. Two numerical methods which compute the fire line evolution have  been developed. 

The first method is exclusively used for single wildfire scenarios and involves finding a numerical approximation of the conformal map from the fire line exterior  to the exterior of the  fixed unit disk in an auxiliary plane. The problem is reduced to a PG-type equation for the unknown conformal map $f$, which, when expressed as a truncated Laurent series, reduces  to a system of $n$ ODEs for the unknown coefficients. The second method is capable of computing the evolution of $J$ wildfires and calculates the pyrogenic potential $\phi$ directly in the physical plane by means of the AAA-LS algorithm \citep{costa2023aaa}. This involves a rational approximation of the harmonic potential involving singularities in the (unphysical) burnt region, which can then be solved by means of a simple least squares method for given boundary data. Unlike the conformal mapping method, the AAA-LS algorithm is explicitly employed at each time step and so an RK time stepping method is used to compute the evolution of the wildfire. Additionally, the multiple wildfire method can handle fire mergers, with the MATLAB \textit{union} function used to combine two overlapping fire lines into one, connected curve.

To demonstrate the efficacy of the numerical models, a number of wildfire scenarios have been simulated. Comparison with previous methods and experimental results \citep{hiltonpyro, sullivan2019investigation} show the ability of the models to produce accurate results efficiently: the quickest with a runtime of seconds and the longest running in under one hour on a standard laptop. With the numerous options available to the user to sacrifice some amount of accuracy for speed, such as decreasing the Laurent series truncation in the conformal mapping method or reducing the time step for a higher order RK method in the multiple spotfires method, there is merit for both methods to be suitable for operational use. All codes used in this work have therefore been made publicly available, see the appropriate Github link.

The effects of the two parameters, the ratio $\alpha$ of the radiative to convective basic ROS effects and the strength $\beta$ of the pyrogenic wind effect, have also been studied in this work. In agreement with the conclusions of \cite{hiltonpyro}, a higher $\beta$ value causes a stronger rounding of the fire head that progresses at a slower rate. Increasing $\alpha$ decreases the ability of two wildfires to grow towards each other which, in the case of zero ambient wind, results in the two extreme limits: $\alpha=0$ where wildfire growth is completely halted on the non-facing portions of the fire line and $\alpha=1$ where there is no interaction between the two wildfires at all. Firebreaks - regions of the fuelbed where wildfire spread is entirely stopped - are straightforward to include into the model, which enables the modelling of wildfires encountering features such as roads, lakes and disconnected man-made fire barriers.

The model \eqref{Meq:rcopa} - \eqref{Meq:far} used in this work is closely related to that used in \cite{hiltonpyro}. Both treat the surface fire as a two-dimensional free boundary problem and model the exterior, pyrogenic wind flow by a Laplace equation of the velocity potential. Further, both consider the same effects on the wildfire spread: a basic ROS, pyrogenic and ambient winds, with equation \eqref{Meq:rcopa} being an exact analogy with equation (12) from \cite{hiltonpyro}. However, there are some noteworthy differences in this work. First is the explicit distinction between radiative and convective rate of spread which in \cite{hiltonpyro} was taken to be a single joint constant $u_0$. Second, the dynamics of the fire plume are treated differently. In the work of \cite{hiltonpyro}, the authors solve an additional Poisson equation in the wildfire interior involving a forcing term $\nu$ representing the upward air flow, whereas the fire plume is observed as a point sink of strength $Q$ in the far field in this work. As shown in section \ref{sec:hcomp}, both of these treatments of the fire plume produce equivalent results that compare well with experimental data. 

Finally, while \cite{hiltonpyro} use a level set method to obtain numerical results, this work uses a conformal mapping method for the single fire scenario and an approach based on the AAA-least squares method when considering multiple wildfire spread and merger. The main purpose of this work is to introduce both of these methods as practical alternatives in wildfire modelling. It is envisioned that the methods detailed here can address some of the shortfalls in existing wildfires tools, specifically modelling unsteady fire propagation dependent on complex wildfire-atmosphere interaction in runtimes suitable for operational use. The methods developed here have further scope for improvement, advancement and extension - some possible extensions are now discussed.

Topography and terrain effects have been ignored in this work yet can play as crucial a role in wildfire spread as the effect of wind. The influence of terrain can be included simply into the wildfire spread equation \eqref{Meq:vn_m} but care is needed in how slopes and valleys would affect the exterior wind field. As suggested in \cite{sharples2020modeling}, an analogue to terrain could be the inclusion of vortices which reproduce terrain effects on the wildfire spread such as vorticity-driven lateral spread. The inclusion of sources, sinks and point vortices into the exterior wind field could additionally be used to model other wildfire features such as fire whirls and tornadoes \citep{soma1991reconstruction,forthofer2011review,tohidi2018fire,lareau2022fire}. As stated in section \ref{sec:fbreak}, inhomogeneous fuel beds could also be incorporated into the model in a similar manner to the firebreak wildfire equation \eqref{Meq:vn_tau}; it would then be simple to couple some existing fuelbed database into the model. Other factors could also be included in equation \eqref{Meq:vn_m}, for example how oxygen availability and transport affects the wildfire spread \citep{zik2,harris2022fingering}.

Further, the merging algorithm could be adapted to allow for more complicated merging scenarios. For example, multiple wildfires may merge in such a way that produces a ring fire - see figure 4.6 in \cite{sharples2022fire} - which results in two disconnected ``inner and outer'' unburnt domains. Locally, the opposite scenario may also appear where a `doughnut' of unburnt fuel with a small fire at its centre is surrounded by a large outer fire on all sides. The AAA-LS algorithm can be extended to account for these types of domains \citep{trefethen2020numerical,costa2023aaa}. In addition, further consideration can be given to the plume strength of each wildfire and of a merged fire. In this work, it is assumed that each fire plume is of equal strength; following a merger, the new wildfire also generates a plume of the same strength as the previous two fires. A mechanism could be included which calculates the plume strength of a merged wildfire given the strengths of its constituent fires and which provides a time varying plume strength related to the size of the wildfire. 

Finally, the current model is entirely deterministic yet wildfire spread, particularly in extreme weather conditions, is often stochastic. Thus stochastic variables could be incorporated into the model, for example an ambient wind with time-varying direction and magnitude or some variability on the basic ROS terms corresponding to stochastic heat fluxes. Another  stochastic wildfire feature is the production and distribution of spotfires as a result of firebrand ejection from a main fire. In this work, the production of spotfires was not considered, rather it was assumed they were present as mature, independent wildfires. The model of this work could be coupled with a stochastic spotfire generation model, for example similar to that in \cite{boychuk2009stochastic}, such that spotfires could appear midway through a scenario and affect the wildfire spread thereafter.

\section*{Acknowledgments}
The authors thank James Hilton, Jason Sharples, Andrew Sullivan, William Swedosh and their team for allowing the use of both their experimental fire data from the CSIRO Pyrotron and the simulation outputs from the pyrogenic potential model in this work. Please see their associated works \cite{hiltonpyro} and \cite{sullivan2019investigation}. Samuel J. Harris was supported by a UK Engineering and Physical Sciences Research Council PhD studentship, Grant No. EP/N509577/1 and EP/T517793/1.

\section*{Software and data availability}
The codes that were used, and the resulting data produced, in this work for calculating the wildfire spread and spotfire merger were created using the MATLAB programming language (version R2024a) and can be found in the following Github repository: \url{https://github.com/Sam-J-Harris/wildfire_spread_spotfire_merger} . This repository was created by Samuel Harris (e-mail: \url{sam.harris.16@ucl.ac.uk}) and contains all MATLAB code used (8.26MB). Author's experimental environment was as follows:
\begin{itemize}
    \item OS: Windows 11
    \item CPU: 11th Gen Intel(R) Core(TM) i7-1185G7 @ 3.00GHz
    \item RAM: 32.00 GB
    \item GPU: Intel(R) Iris(R) Xe Graphics
\end{itemize}
The data produced in this work is also made publicly available at the above Github link. Data for all figures can be found in the ``00 Data'' folder in the form of .mat (boundary data for the evolving fire lines) and .fig (images of the fire line evolution) files. The only other data used in this work were images of the experimental fires detailed in \cite{sullivan2019investigation} and the numerical outputs of \cite{hiltonpyro}. This data was provided to us by the authors of these papers. 

\bibliographystyle{apacite} 
\bibliography{ABrefs.bib}

\end{document}